\documentclass[aps,prb,superscriptaddress,reprint]{revtex4-2}
\usepackage{graphicx}
\usepackage{amsmath,amssymb}
\usepackage{bm}
\usepackage[T1]{fontenc}
\usepackage{lmodern}
\usepackage{hyperref}
\usepackage{xcolor}
\usepackage{empheq}
\hypersetup{colorlinks=true,citecolor={blue},linkcolor={blue},urlcolor={blue}}
\usepackage{soul}

\begin{document}

\title{Theory for Coulomb scattering of trions in 2D materials}

\author{Kok Wee Song}
\affiliation{Department of Physics and Astronomy, University of Exeter, Stocker Road, Exeter EX4 4QL, United Kingdom}

\author{Salvatore Chiavazzo}
\affiliation{Department of Physics and Astronomy, University of Exeter, Stocker Road, Exeter EX4 4QL, United Kingdom}

\author{Ivan A. Shelykh}
\affiliation{Science Institute, University of Iceland, Dunhagi 3, IS-107, Reykjavik, Iceland}
\affiliation{Department of Physics and Engineering, ITMO University, St. Petersburg 197101, Russia}

\author{Oleksandr Kyriienko}
\affiliation{Department of Physics and Astronomy, University of Exeter, Stocker Road, Exeter EX4 4QL, United Kingdom}

\date{\today}

\begin{abstract}
We develop a theoretical description of Coulomb interactions between trions (charged excitons) that define a nonlinear optical response in doped two-dimensional semiconductors. First, we formulate a microscopic theory of trion-trion interactions based on composite nature of these particles, and account for all possible exchange processes. Next, we calculate numerically the trion binding energies and corresponding three-body wavefunctions using a basis set with high expressivity. Then, using the obtained wavefunctions we calculate the matrix elements of two-trion scattering, and compare the contributions coming from direct and exchange terms. Finally, we find that the considered scattering gives significant contribution to the optical nonlinearity in monolayers of transition metal dichalcogenides. In particular, this can lead to an attractive interaction in doped monolayers. Our theory opens a route for studying nonlinear properties of trion-polaritons inaccessible before.
\end{abstract}

\maketitle

\section{Introduction}

An optical response of low-dimensional direct bandgap semiconductors is often defined by excitons---bound electron-hole (e-h) pairs \cite{HaugKoch2004}. Arising due to the Coulomb attraction, excitons have a finite volume and effectively enhance light-matter coupling \cite{Weisbuch:PRL1992}. When embedded into high-Q optical microcavities, excitons can enjoy strong light-matter coupling (SC) \cite{Basov2021}. Hybridizing with cavity photons, they become exciton-polaritons \cite{LIEW20111543,Sanvitto2016}. One prominent example of polaritonic platform is III-V semiconductor quantum wells (e.g. GaAs/AlGaAs), where SC was achieved~\cite{Kasprzak2006,Ferrier2011,Walker2013,Cilibrizzi2014,Walker2015,Rodriguez2016b}. Thanks to exciton-exciton interactions, polaritons gain nonlinearity, and to date various nonlinear optical effects were observed at cryogenic temperatures in the regime of macroscopic occupation~\cite{Hivet2012,Abbarchi2013,Walker2017b,Sich2018,Topfer:21,Alyatkin2021,Baryshev2022}. Recent efforts were directed towards enhancing interactions with the goal of developing a quantum polaritonic platform~\cite{Liew:PRL2010,Kyriienko2014a,Kyriienko2014b,Kyriienko2016,Flayac2017,Snijders:PRL2018}, and nonlinear polariton effects are getting studied at few-photon level \cite{Munoz-Matutano2019,Delteil2019,Cuevas2018,Togan:PRL2018,Zasedatelev2021,Kuriakose2022}. However, the value of the nonlinear interaction constant is ultimately limited by material properties, and the way excitons interact. Namely, the electrons and holes constituting two excitons (composite bosons) can exchange and interact via Coulomb potential \cite{Combescot:PhysRep463(2008),Shiau:EPL117(2017)}, leading to a major short-range scattering contribution \cite{Ciuti:PRB1998,Tassone1999}. Another nonlinear process comes from a phase-space filling and the related nonlinear saturation \cite{Tassone1999,Schmitt-Rink1985,Brichkin2011,Emmanuele:NatCommun11(2020)} and corrections beyond the Born approximation~\cite{Levinsen:PRR1(2019),Bleu2020}. The value of nonlinearity is mainly defined by a ground state wavefunction, which is either known in the analytical form~\cite{HaugKoch2004} or can be approximated to high precision~\cite{Kidd:PRB93(2016)}. One finds that the nonlinear response due to Coulomb scattering is simply proportional to the exciton size~\cite{Tassone1999,Shahnazaryan:PRB96(2017)}, where a larger overlap between excitonic wavefunctions leads to a larger scattering cross-section.
\begin{figure}
    \centering
    \includegraphics[width=3.3in]{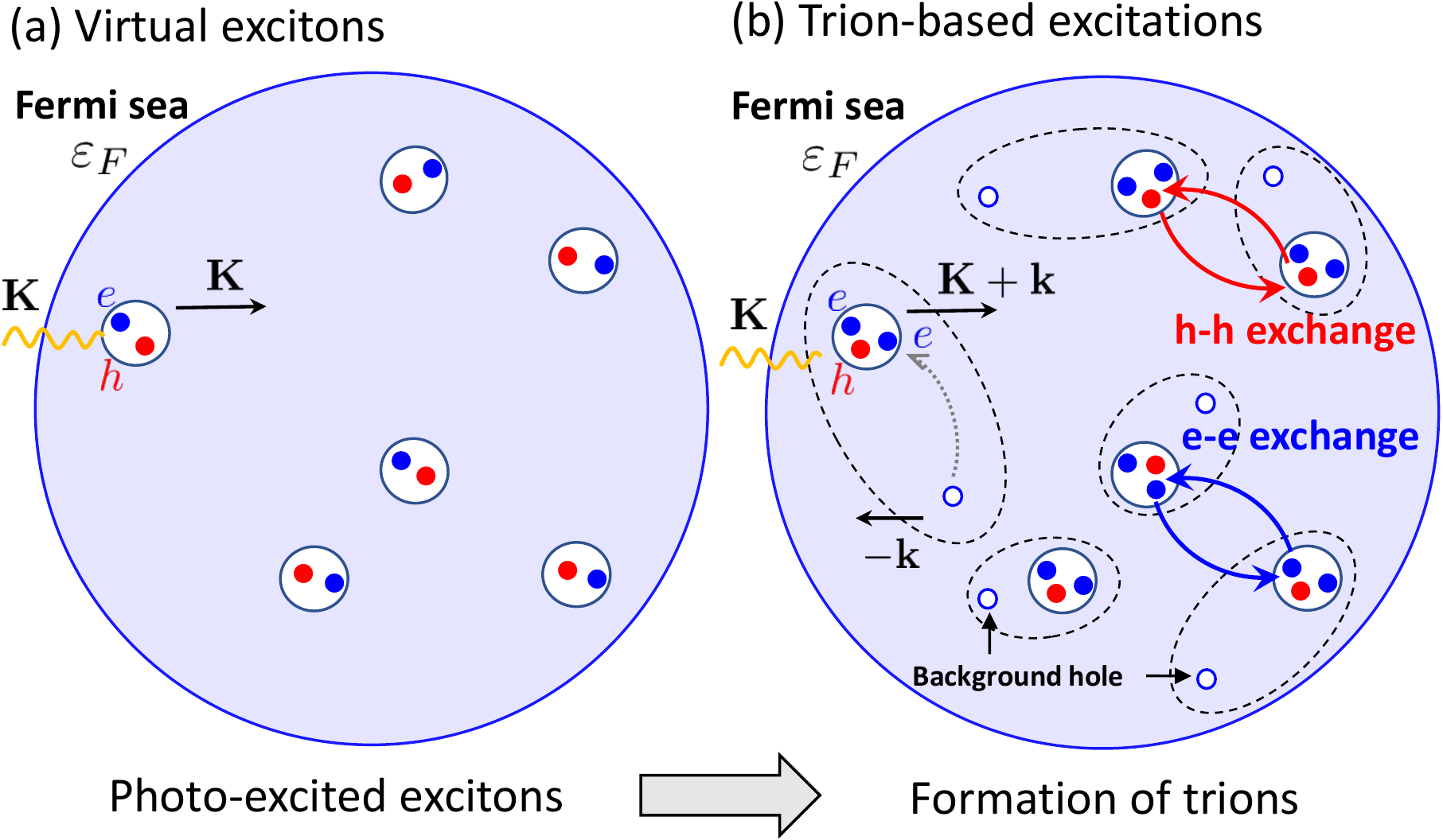}
    \caption{\textbf{Sketch of trion-based excitations.} (a) Excitonic complexes (bound electron-hole pairs) are virtually created by absorbing photons with wavevector $\mathbf{K}$. 
    Here, blue dots depict electrons and red dots depict holes. In the presence of doping $\varepsilon_{\mathrm{F}}$ excitons develop correlations with electrons in the Fermi sea forming a trion (white disk with two blue and one red dot). (b) Photoexcited complexes correspond to trion-hole pairs (black dashed ellipsoids) that include background hole (blue open circles), being the empty state in the Fermi sea. The nonlinear interaction between trion complexes originates from electron-electron (blue arrows) and hole-hole (red arrows) exchange processes, leading to the short-range Coulomb interaction between trions.}
    \label{fig:T-h}
\end{figure}

Two-dimensional (2D) semiconductors, represented by monolayers of transition metal dichalcogenides (TMD), possess excellent optical properties~\cite{Mak:NatMat12(2012),Splendiani:NanoLett10(2010),Wurstbauer2017,Berkelbach2018} due to robust excitons that are dominant even at room temperature~\cite{Chernikov:PRL115(2015),Ugeda:NatMat:13(2014),He:PRL113(2014),Zhang:PRB89(2014),You:NatPhys11(2015),Singh:PRB93(2016),Lundt:APL112(2018),Deilmann:NanoLett18(2018)}. Due to a small localization volume, TMD excitons can enter the SC regime already with a single monolayer \cite{Wang:RMP90(2018),Mueller2018,Kaviraj2019}. This has led to the boom of 2D polaritonics~\cite{Liu:NanoLett16(2016),Sidler:NatPhys13(2016),Dufferwiel:NatPhoto11(2017),Dhara:NPhys14(2017),Cuadra:NanoLett18(2018),Barachati:NatNano13(2018),Emmanuele:NatCommun11(2020),Tan:PRX10(2020),Kravtsov:LightSci9(2020),Stepanov:PRL126(2021),Anton-Solanas2021,Schneider2018}. However, large excitonic binding energy in these materials has its cost: the nonlinearity coming from exciton-exciton (X-X) scattering becomes weakened \cite{Shahnazaryan:PRB96(2017)}, as compared to conventional semiconductor materials such as GaAs \cite{Ciuti:PRB1998,Estrecho2019}, due to the reduction of the exciton size. This determines smaller exciton nonlinearity coming from X-X scattering in TMD materials~\cite{Barachati:NatNano13(2018),Emmanuele:NatCommun11(2020)}, and larger influence of saturation~\cite{Emmanuele:NatCommun11(2020)}.

Luckily, together with neutral excitons, TMD materials at finite doping reveal a presence of quasiparticles based on photoexcited e-h pairs correlated with free carriers. These quasiparticles correspond to exciton-polarons or trion-based excitations \cite{Mak2013,Courtade:PRB96(2017)}. Specifically, trions appear from a photoexcited exciton that is bound to a free electron, leaving an empty state in a Fermi sea (see the sketch in Fig.~\ref{fig:T-h}). First observed in doped GaAs quantum wells \cite{Rapaport:PRL84(2000),Rapaport:PRB63(2001)}, trion-polaritons can enter SC, albeit with smaller values of light-matter coupling and only at low temperatures (limited by small bindings). In TMD materials trion-polaritons emerge as robust quasiparticles at finite doping, as shown in various experiments~\cite{Sidler:NatPhys13(2016),Emmanuele:NatCommun11(2020),Kravtsov:LightSci9(2020)}. Trion-polaritons revealed an increased nonlinearity as they have different nonlinear saturation mechanisms~\cite{Kyriienko:PRL125(2020),Tan:PRX10(2020)} and due to extended size shall exhibit stronger trion-trion (T-T) Coulomb scattering. At the same time, the theoretical description of the full nonlinear optical response for trion-polaritons is still missing. The current paper aims to resolve this.

Construction of the theory for trion-trion scattering is a challenging task~\cite{Shiau:PRB86(2012)}. Even at a single excitation level, a coherent description of a trion and its coupling to light is much more tricky than of an exciton. This is due to the complex correlated nature of a former \cite{Shiau:EPL117(2017),Ravets:PRL120(2018),BastarracheaMagnani:PRL126(2021),Chang:PRB98(2018),Rana:PRB102(2020),Rana:PRL126(2021),Koksal:PRR3(2021),Kyriienko:PRL125(2020),Shahnazaryan:PRB102(2020),Zhumagulov:NPJ2022,Li:PRL126(2021),Li:PRB103(2021),Efimkin:PRB95(2017),Efimkin:PRB103(2021),Katsch2022, Denning:PRR4(2022),Denning:PRB105(2022),Tiene:PRB105(2022)}. The corresponding quantum three-body problem has no analytical solution~\cite{Combescot:PRX7(2017)}, and one is restrained to use either simplified ansatz~\cite{Ramon:PRB67(2003),Courtade:PRB96(2017)} or recur to the direct \textit{ab initio} procedure \cite{Zhumagulov:JChemPhys153(2020),Zhumagulov:PRB101(2020),Zhumagulov:NPJ2022}. Together with a large number of scattering diagrams (108 terms) and the presence of background charge, this poses a challenge for the analysis of the T-T interactions. Similar to X-X interaction, the dominant exchange channel for trions is very sensitive to their wavefunctions. We show to overcome this issue.

In this paper, we developed a theory for composite trion scattering. Using the method of the decomposition of a three body wavefunction with large Gaussian basis set, we obtain an accurate description for trion ground state in 2D materials, and compute the T-T exchange scattering. We also analyze dependence of T-T scattering on materials parameters. In the accompanying letter, we reveal the strong enhancement of trion-polariton nonlinearity as compared to the excitonic case, and find T-T attraction in TMD monolayers \cite{Song2022}.

The paper is organized as follows. The model Hamiltonian is presented in Section \ref{sec:model}. In Section \ref{sec:TTInteraction} we present the matrix elements of T-T interaction, with details of the derivation shown in Appendices. In Section \ref{sec:numerical} we discuss the method used for the numerical calculation of these matrix elements \ref{sec:TTInteraction}. In Section \ref{sec:nonlinearity} we analyze the nonlinear response as a function material parameters, and refer to our accompanying letter \cite{Song2022} for the particular case of TMD monolayer. Conclusions are presented in Section \ref{sec:conclusion}.

\section{Model}\label{sec:model}

In this section, we present a microscopic Hamiltonian for our system and discuss the formation of trions. Then, we describe their interaction with microcavity photon mode that can lead to the formation of trion-polaritons.

\subsection{Single trion state}

We consider a system of two-dimensional semiconductor. We start with a two-band model that describes electrons in a conduction band and holes in a valence band. They interact with each other via Coulomb interaction. The corresponding Hamiltonian reads
\begin{equation}\label{eqn:H}
    \mathcal{H}_e=\sum_{\mathbf{k}}(\varepsilon_\mathbf{k}^ca^\dagger_\mathbf{k}a_{\mathbf{k}}+\varepsilon_\mathbf{k}^vb^\dagger_\mathbf{k}b_{\mathbf{k}})+\frac{1}{2L^2}\sum_{\mathbf{q}}W(\mathbf{q})\rho_{\mathbf{q}}\rho_{-\mathbf{q}},
\end{equation}
where $a_{\mathbf{k}}$ ($a^{\dagger}_{\mathbf{k}}$) and $b_{\mathbf{k}}$ ($b^\dagger_{\mathbf{k}}$) are annihilation (creation) field operators for conduction and valence bands with the in-plane momentum $\mathbf{k}$. $\varepsilon^{c,v}_{\mathbf{k}}$ are the conduction and valence band dispersions and $W(\mathbf{q})$ is the screen intra and interband Coulomb interacting potential. We also introduce the total density operator as  $\rho_{\mathbf{q}}=\sum_{\mathbf{k}}(a^\dagger_{\mathbf{k}+\mathbf{q}}a_{\mathbf{k}}+b^\dagger_{\mathbf{k}+\mathbf{q}}b_{\mathbf{k}})$. 

We introduce a trion creation operator as a composite three-body operator that is composed by two electrons and one hole,
\begin{align}\label{eqn:trion}
\mathcal{T}^{\dagger}_{\alpha\mathbf{Q}}
=
\sum_{\mathbf{k}_1\mathbf{k}_2}\frac{1}{\sqrt{2!}}A^{\alpha\mathbf{Q}}_{\mathbf{k}_1\mathbf{k}_2}a^\dagger_{\mathbf{k}_1}a^\dagger_{\mathbf{k}_2}b_{\mathbf{k}_1+\mathbf{k}_2-\mathbf{Q}},,
\end{align}
where $\alpha$ labels a trion state with momentum $\mathbf{Q}$. The coefficients  $A^{\alpha\mathbf{Q}}_{\mathbf{k}_1\mathbf{k}_2}=-A^{\alpha\mathbf{Q}}_{\mathbf{k}_2\mathbf{k}_1}$ are anti-symmetric with respect to the interchange of $\mathbf{k}_1$ and $\mathbf{k}_2$. 

The trion wavefunction in Eq.~\eqref{eqn:trion} satisfies the three-body Wannier equation \cite{Combescot:PhysRep463(2008),Deilmann:PRL116(2016),Drueppel:NatComm8(2017),Torche:PRB100(2019),Zhumagulov:PRB101(2020),Zhumagulov:JChemPhys153(2020)}
\begin{align}\label{eqn:T-WannierQ}
(\varepsilon^{c}_{\mathbf{k}_1}\!+\!\varepsilon^{c}_{\mathbf{k}_2}\!-\!\varepsilon^{v}_{\mathbf{k}_1+\mathbf{k}_2-\mathbf{Q}})A^{\alpha \mathbf{Q}}_{\mathbf{k}_1\mathbf{k}_2}
+&\frac{1}{L^2}\sum_{{\mathbf{q}_1\mathbf{q}_2}}\Xi_{\mathbf{q}_1\mathbf{q}_2}A^{\alpha\mathbf{Q}}_{\mathbf{k}_1-\mathbf{q}_1,\mathbf{k}_2-\mathbf{q}_2}\notag\\
&
=
E_{\alpha\mathbf{Q}}A^{\alpha \mathbf{Q}}_{\mathbf{k}_1\mathbf{k}_2},
\end{align}
where the interaction kernel is 
$\Xi_{\mathbf{q}_1\mathbf{q}_2}=W(\mathbf{q}_2)\delta_{\mathbf{q}_1,-\mathbf{q}_2}-W(\mathbf{q}_1)\delta_{\mathbf{q}_2,0}-W(\mathbf{q}_2)\delta_{\mathbf{q}_1,0}$, and $L^2$ denotes an area of the system. 

\subsection{$N$-trion state}

To create trions optically, one naturally needs to have a vacuum state that corresponds to the presence of free electrons in the conduction band, 
\begin{equation}
|\varnothing\rangle=\prod_{\varepsilon^c_{\mathbf{k}}\leq \varepsilon_{\mathrm{F}}}a^\dagger_{\mathbf{k}}|0\rangle , 
\end{equation}
where $|0\rangle$ is a ground state in the absence of doping. In this case, photoexcitation of a trion can be viewed as creating a neutral exciton, which then captures an additional electron from the Fermi sea (see Fig.~\ref{fig:T-h}). This can also be seen as a formation of exciton-polaronic quasiparticle in the regime of weak doping \cite{Glazov:JChemPhys153(2020),Efimkin:PRB103(2021),Efimkin:PRB95(2017),Li:PRL126(2021),Li:PRB103(2021)}.  This four-body excitation can be represented as a linear combination of all pairs of a trion with a hole in the Fermi sea satisfying the condition of momentum conservation \cite{Rapaport:PRL84(2000),Rapaport:PRB63(2001),Rana:PRB102(2020),Rana:PRL126(2021),Koksal:PRR3(2021)}. Following the model developed in Refs.~\cite{Rapaport:PRL84(2000),Rapaport:PRB63(2001)} we write the operator of the trion-hole excitation (T-h) as
\begin{equation}\label{eqn:B}
    \mathcal{B}^\dagger_{\alpha \mathbf{K}}=\frac{1}{\sqrt{N_{\mathrm{F}}}}\sum_{|\mathbf{k}|\leq k_{\mathrm{F}}}\mathcal{T}^\dagger_{\alpha,\mathbf{K}+\mathbf{k}}a_\mathbf{k},
\end{equation}
where $N_{\mathrm{F}}$ is a number of electrons in the Fermi sea.
The state of $N$ trions can be constructed as
\begin{equation}
    |N\rangle=\frac{(\mathcal{B}^\dagger_{\alpha \mathbf{k}})^{N}}{\sqrt{N!F_{N}}}|\varnothing\rangle,
\end{equation}
where
\begin{equation}
F_{N}=\frac{\langle \varnothing|(\mathcal{B}_{\alpha \mathbf{k}})^{N-1}(\mathcal{B}^\dagger_{\alpha \mathbf{k}})^{N-1}|\varnothing\rangle}{N!}    
\end{equation}
is a normalization constant that accounts for the composite nature of quasiparticles.

\section{Matrix elements of trion-trion interaction}\label{sec:TTInteraction}

We proceed considering energy corrections to the $N$-trion energy arising due to Coulomb potential. In the presence of T-T interactions the total energy of the $N$-trion state is no longer linearly proportional to $N$, the total number of trions. This corresponds to the appearance of a quartic interaction term, and corresponding Kerr-type nonlinearity for trion-polaritons. The strength of this nonlinearity can be characterized as an energy of creating an additional trion in the $N$-trion system,
\begin{equation}
    \Delta^{(N)}_{\alpha\mathbf{K}}=\langle N+1|\mathcal{H}_e|N+1\rangle-\langle N|\mathcal{H}_e|N\rangle.
\end{equation}

To evaluate $\Delta^{(N)}_{\alpha\mathbf{K}}$ we use the composite particle approach ~\cite{Combescot:PhysRep463(2008),Combescot:EurPhysJB79(2011),Glazov2009}. It relies on the commutation relations for the composite field operators $\mathcal{T}^\dagger_{\alpha\mathbf{K}}$ and $\mathcal{T}_{\alpha\mathbf{K}}$, which do not obey neither bosonic nor fermionic algebra (see Appendix~\ref{app:commutator}). In this section, we outline the main steps required for the quantitative evaluation of the trion-polariton nonlinearity. A more detailed discussion is presented in Appendix~\ref{app:nonlinearity}.

To calculate the $N$-trion energy, we first show that
\begin{equation}\label{eqn:[H,T]}
[\mathcal{H}_e,\mathcal{T}^\dagger_{\alpha\mathbf{Q}}]\!=\!E_{\alpha\mathbf{Q}}\mathcal{T}^\dagger_{\alpha\mathbf{Q}}\!+\!\!\sum_{\alpha'\mathbf{q}}\!\frac{W(\mathbf{q})}{L^2}F^{\alpha\alpha'}_{\mathbf{Q},\mathbf{Q}+\mathbf{q}}\mathcal{T}^{\dagger}_{\alpha',\mathbf{Q}+\mathbf{q}}\rho_{-\mathbf{q}},
\end{equation}
where 
\begin{equation}
F^{\alpha\alpha'}_{\mathbf{Q},\mathbf{Q}+\mathbf{q}}\!\!=\!\!\sum\limits_{\mathbf{k}_1\mathbf{k}_2}[A^{\alpha\mathbf{Q}}_{\mathbf{k}_1-\mathbf{q},\mathbf{k}_2}\!+\!A^{\alpha\mathbf{Q}}_{\mathbf{k}_1,\mathbf{k}_2-\mathbf{q}}\!\!-A^{\alpha \mathbf{Q}}_{\mathbf{k}_1\mathbf{k}_2}] (A^{\alpha',\mathbf{Q}-\mathbf{q}}_{\mathbf{k}_1\mathbf{k}_2})^\ast
\end{equation}
is a wavefunction overlap factor for the Coulomb scattering of two trions.

Using Eq.~\eqref{eqn:[H,T]}, we derive
\begin{align}
    [\mathcal{H}_e,\mathcal{B}^\dagger_{\alpha\mathbf{K}}]\approx&
    E_{\alpha\mathbf{K}}\mathcal{B}^\dagger_{\alpha\mathbf{K}}
    +\frac{1}{L^2}\sum_{\mathbf{q}}W(\mathbf{q})\mathcal{C}^\dagger_{\alpha\mathbf{K}}(\mathbf{q})\rho_{-\mathbf{q}},\label{eqn:[H,B]}
\end{align}
where we have let $E_{\alpha,\mathbf{K}+\mathbf{k}}\approx E_{\alpha\mathbf{K}}$ be a slow-varying function of the total momentum. In Eq.~\eqref{eqn:[H,B]} we introduced operators
\begin{align}
\mathcal{C}^\dagger_{\alpha,\mathbf{K}}(\mathbf{q})=&\frac{1}{\sqrt{N_{\mathrm{F}}}}\sum_{\alpha',\mathbf{k}}G^{\alpha,\alpha'}_{\mathbf{K}}(\mathbf{k},\mathbf{q})\mathcal{T}^\dagger_{\alpha',\tilde{\mathbf{k}}+\mathbf{q}}a_\mathbf{k},\\
G^{\alpha\alpha'}_{\mathbf{K}}(\mathbf{k},\mathbf{q})=&F^{\alpha\alpha'}_{\tilde{\mathbf{k}},\tilde{\mathbf{k}}+\mathbf{q}}\Theta_{k_{\mathrm{F}}}(|\mathbf{k}|)-\delta_{\alpha\alpha'}\Theta_{k_{\mathrm{F}}}(|\mathbf{k}+\mathbf{q}|),
\end{align}
where $\tilde{\mathbf{k}}:=\mathbf{K}+\mathbf{k}$ is defined to simplify expressions. $\Theta_{k_{\mathrm{F}}}(|\mathbf{k}|)=1$ for $|\mathbf{k}|\leq k_{\mathrm{F}}$ and zero otherwise (namely, the Heaviside step function).

\begin{figure*}
    \centering
    \includegraphics[width=5.3in]{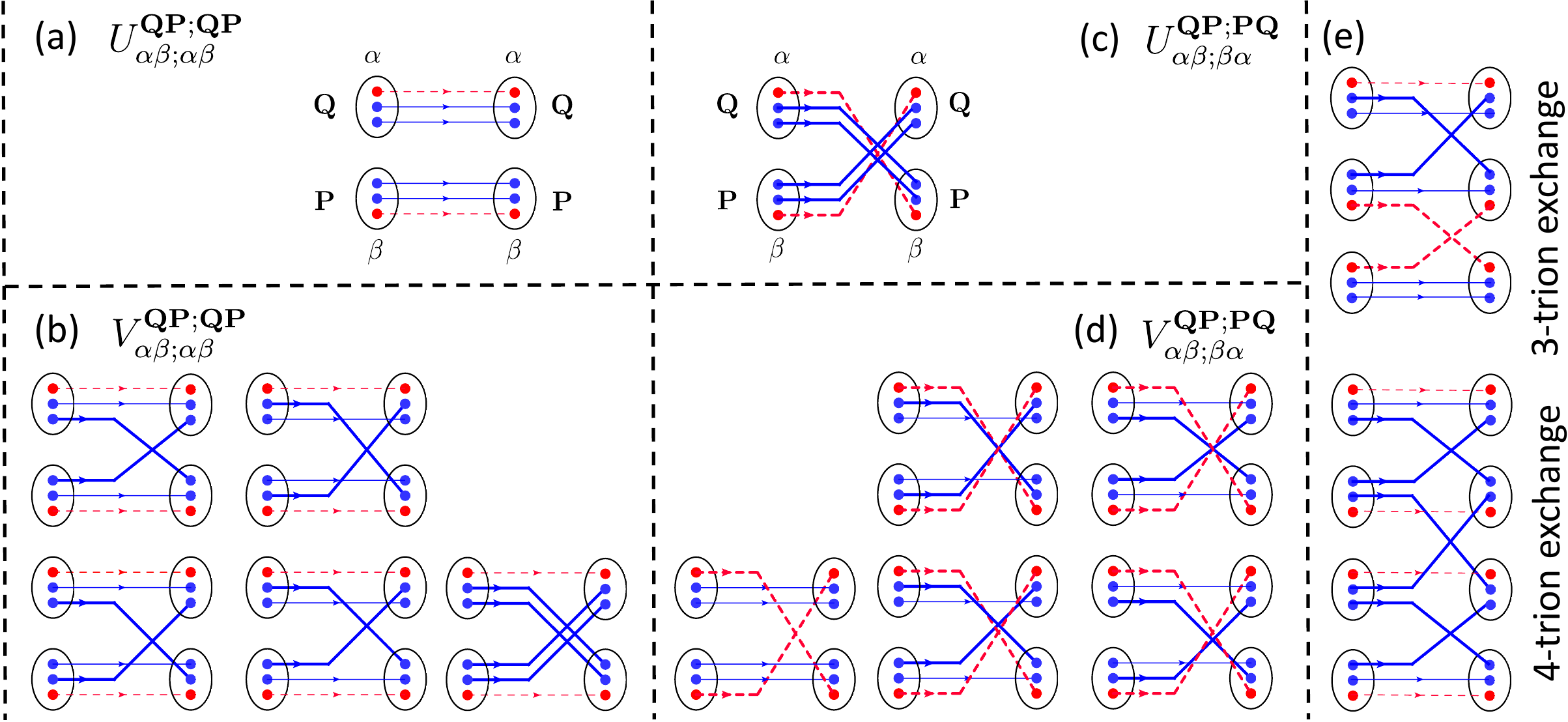}
    \caption{\textbf{Scattering diagrams for composite particles} (also known as Shiva diagrams) that illustrate exchanges of elementary particles (electrons and holes) during the trion-trion scattering processes [Eqs.~\eqref{eqn:Ubar} and \eqref{eqn:Vbar}]. A trion bound state is depicted by an ellipsoid with two blue dots (electrons) and one red dot (hole). $\alpha$ and $\beta$ label the state of the trions. 
    The lowest-order processes (2-trion exchange) include: (a) direct T-T interaction with no particles exchanged; (b) leading nonzero exchange process giving rise to the nontrivial effects due to the composite properties of trion; (c) exchange process that treats trion as an elementary fermion. In panel (d) we depict processes equivalent to those in (b) but with an additional exchange of the two trions [three particles exchange, like (c)] in the final state. Panel (e) we present examples of higher-order exchange processes.}
    \label{fig:shiva}
\end{figure*}

Using Eq.~\eqref{eqn:[H,B]} and the condition
$b^\dagger_{\mathbf{k}}|\varnothing\rangle=0$ (as the valence band is fully filled in the vacuum state), we arrive at
\begin{align}
     &\langle N|\mathcal{H}_e|N\rangle= NE_{\alpha\mathbf{K}}+\frac{N(N-1)/N_{\mathrm{F}}}{L^2\sqrt{(N-2)!F_{N-2}}}\frac{F_{N-2}}{4F_N}\notag\\
     &\sum_{\mathbf{q}}W(\mathbf{q})\langle\varnothing|
    \mathcal{B}^2_{\alpha\mathbf{K}}\mathcal{C}^\dagger_{\alpha\mathbf{K}}(-\mathbf{q})\mathcal{C}^\dagger_{\alpha\mathbf{K}}(\mathbf{q})
    \mathcal{B}_{\alpha\mathbf{K}}^{N-2}| N-2\rangle,\label{eqn:<N|H|N>}
\end{align}
if only the two-trion exchange processes are accounted for (see Appendix~\ref{app:nonlinearity} for the full derivation). We note that higher-order exchange processes, such as simultaneous 3- and 4-trions scattering [Fig.~\ref{fig:shiva}(e)], are suppressed in the low-density limit \cite{Combescot:PhysRep463(2008)}.

We proceed by using the anticommutation relation (Appendix \ref{app:commutator}) for trion operators
\begin{equation}
 \{\mathcal{T}_{\alpha'\mathbf{Q}'},\mathcal{T}^\dagger_{\alpha\mathbf{Q}}\}=\delta_{\alpha'\alpha}\delta_{\mathbf{Q},\mathbf{Q}}+\mathcal{D}^{\alpha'\alpha}_{\mathbf{Q}'\mathbf{Q}}    
\end{equation}
and move all trion creation operators to the left, and use the fact that $\langle\varnothing|\mathcal{T}^\dagger_{\alpha\mathbf{Q}}=0$. As compared to elementary fermions, the anticommutator for trion fields has an extra operator $\mathcal{D}$, arising from their composite nature. To move completely  $\mathcal{T}^{\dagger}_{\alpha\mathbf{Q}}$ to the right in Eq.~\eqref{eqn:<N|H|N>}, one needs to use an additional commutation relation 
\begin{align}
     [\mathcal{D}^{\alpha'\alpha}_{\mathbf{Q}'\mathbf{Q}},\mathcal{T}^\dagger_{\beta\mathbf{P}}]=&\sum_{\bar{\alpha} \bar{\mathbf{Q}}}\mathcal{T}^\dagger_{\bar{\alpha}\bar{\mathbf{Q}}}\Big[\Big(\Lambda^{\alpha\beta,\alpha'\bar{\alpha}}_{\mathbf{Q}\mathbf{P},\mathbf{Q}'\bar{\mathbf{Q}}}-\Lambda^{\alpha\beta,\bar{\alpha}\alpha'}_{\mathbf{Q}\mathbf{P},\bar{\mathbf{Q}}\mathbf{Q}'}\Big)\notag\\
    &\times\delta_{\bar{\mathbf{Q}},\mathbf{Q}-\mathbf{Q}'+\mathbf{P}}+\mathcal{P}^{\alpha\beta,\alpha'\bar{\alpha}}_{\mathbf{Q}\mathbf{P},\mathbf{Q}'\bar{\mathbf{Q}}}\Big],\label{eqn:[D,T]}
\end{align}
where
\begin{align}
    &\Lambda^{\alpha\beta,\alpha'\beta'}_{\mathbf{Q}\mathbf{P},\mathbf{Q}'\mathbf{P}'}\!=\!\sum_{\mathbf{k}_1\mathbf{k}_2}\sum_{\mathbf{p}_1\mathbf{p}_2}\Big(A^{\alpha'\mathbf{Q}'\ast}_{\mathbf{p}_1\mathbf{p}_2}\!A^{\beta'\mathbf{P}'\ast}_{\mathbf{k}_1\mathbf{k}_2}
    \delta_{\mathbf{p}_1+\mathbf{p}_2-\mathbf{Q}',\mathbf{k}_1+\mathbf{k}_2-\mathbf{Q}}
    \notag\\
    &
    -4A^{\alpha'\mathbf{Q}'\ast}_{\mathbf{k}_2\mathbf{p}_2}\!A^{\beta'\mathbf{P}'\ast}_{\mathbf{k}_1,\mathbf{p}_1}\delta_{\mathbf{p}_2-\mathbf{Q}',\mathbf{k}_1-\mathbf{Q}}
    \Big)A^{\alpha\mathbf{Q}}_{\mathbf{k}_1\mathbf{k}_2}\!A^{\beta\mathbf{P}}_{\mathbf{p}_1\mathbf{p}_2}
    \label{eqn:Lambda}
\end{align}
is the Pauli scattering term~\cite{Combescot:PhysRep463(2008),Glazov2009} which encodes all the essential information for 2-trion wavefunction overlapping in the exchange processes.
$\mathcal{P}$ is a two-body operator (Appendix \ref{app:commutator}) containing $a^\dagger_{\mathbf{k}} a_{\mathbf{k}'}$ and $b^\dagger_{\mathbf{k}} b_{\mathbf{k}'}$. We discard this term in our further consideration, since it corresponds to the 3-trion scattering \cite{Combescot:PRL104(2010)}.  

Under a typical condition $\mathcal{D}^{\alpha'\alpha}_{\mathbf{Q}\mathbf{Q}'}|\varnothing\rangle=0$ \cite{Combescot:PhysRep463(2008)}, one gets
\begin{align}
     \langle N|\mathcal{H}_e|N\rangle&\approx NE_{\alpha\mathbf{K}}+\frac{N(N-1)}{2L^2}\frac{F_{N-2}}{F_N}g_{\mathrm{T}} ,
\end{align}
where $g_{\mathrm{T}}$ is the Kerr nonlinearity coefficient,
\begin{align}
    g_{\mathrm{T}}\!=\!
    \!{\sum_{\mathbf{p}_1\mathbf{p}_2}}'
    \Big[
    \!-\!\bar{U}_{\alpha\alpha,\alpha\alpha}^{\tilde{\mathbf{p}}_1\tilde{\mathbf{p}}_2,\tilde{\mathbf{p}}_2\tilde{\mathbf{p}}_1}\!+\!\bar{V}_{\alpha\alpha,\alpha\alpha}^{\tilde{\mathbf{p}}_1\tilde{\mathbf{p}}_2,\tilde{\mathbf{p}}_1\tilde{\mathbf{p}}_2}\!-\!\bar{V}_{\alpha\alpha,\alpha\alpha}^{\tilde{\mathbf{p}}_1\tilde{\mathbf{p}}_2,\tilde{\mathbf{p}}_2\tilde{\mathbf{p}}_1}\Big]
    \label{eqn:gT}
\end{align}
and ${\sum_{\mathbf{k}}}'\equiv\sum_{|\mathbf{k}|\leq k_{\mathrm{F}}}$.
In the above equation,
\begin{equation}\label{eqn:Ubar}
    \bar{U}_{\alpha \beta\alpha' \beta'}^{\tilde{\mathbf{p}}_1\tilde{\mathbf{p}}_2,\tilde{\mathbf{p}}_1-\mathbf{q},\tilde{\mathbf{p}}_2+\mathbf{q}}\!=\!G^{\alpha\alpha'}_{\mathbf{K}}(\mathbf{p}_1,\!-\mathbf{q})W(\mathbf{q})G^{\beta\beta'}_{\mathbf{K}}(\mathbf{p}_2,\mathbf{q}),
\end{equation}
is the renormalized direct interaction (subtracted by the background charges) and
\begin{equation}\label{eqn:Vbar}
    \bar{V}_{\alpha\beta,\alpha'\beta'}^{\tilde{\mathbf{p}}_1\tilde{\mathbf{p}}_2,\tilde{\mathbf{p}}_1'\tilde{\mathbf{p}}_2'}\!=\!\sum_{\mu\nu\mathbf{q}}\!\bar{U}_{\alpha\beta,\mu\nu}^{\tilde{\mathbf{p}}_1\tilde{\mathbf{p}}_2,\tilde{\mathbf{p}}_1-\mathbf{q},\tilde{\mathbf{p}}_2+\mathbf{q}}
    \Lambda^{\mu\nu,\alpha'\beta'}_{\tilde{\mathbf{p}}_1-\mathbf{q},\tilde{\mathbf{p}}_2+\mathbf{q};\tilde{\mathbf{p}}_1',\tilde{\mathbf{p}}_2'},
\end{equation}
is the renormalized exchange interaction. In Fig.~\ref{fig:shiva}(a)-(d) we illustrate
all possible exchange processes for the T-T scattering (without the subtraction of background charges). 

When calculating the nonlinearity coefficient, it is important to account in the direct term of Eq.~\eqref{eqn:gT} the effect of the compensation of the long-range Coulomb repulsion tail. Indeed, the nonlinear response only includes the additional energy of the interaction of two trions with respect to the energy of two bare interacting electrons, and the latter has to be subtracted. Therefore, instead of the standard $1/q$ Coulomb divergence at $q\rightarrow0$, one gets $\bar{U}_{\alpha \alpha;\alpha \alpha}^{\tilde{\mathbf{p}}_1\tilde{\mathbf{p}}_2;\tilde{\mathbf{p}}_1,\tilde{\mathbf{p}}_2}=0$.  The procedure is similar to the one employed for the jellium model~\cite{Mahan:Many-Particle(2000)}, where the divergent direct interaction term is compensated by the positive background.

In the large-$N$ limit, we can approximate the statistical factor as $F_{N-2}/F_{N} \approx 1$ \cite{Combescot:PhysRep463(2008)} and get
\begin{equation}
    \Delta^{(N)}_{\alpha\mathbf{K}}\approx E_{\alpha \mathbf{K}}+n_{\mathrm{T}} g_{\mathrm{T}},
\end{equation}
where $n_{\mathrm{T}}=N/L^2$ is the trion density. For the generic trion wavefunction, it is a challenging task to evaluate $g_{\mathrm{T}}$ as there does not exist an analytical expression that allows estimating it faithfully \cite{Combescot:PRX7(2017)}. To solve this problem, we assume that the trion wavefunction is a slow-varying function of its total momentum, and keep only the leading exchange processes which are depicted by the exchange (``Shiva'') diagrams in Fig.~\ref{fig:shiva}(c). 

We then obtain the main result of this paper, being the T-T interaction constant
\begin{align}
    &g_{\mathrm{T}}\approx\sum_{\mathbf{q}}
    \sum_{\mathbf{k}_1\mathbf{k}_2}\sum_{\mathbf{k}_1'\mathbf{k}_2'}W(\mathbf{q})
    \Gamma^\alpha_{\mathbf{k}_1\mathbf{k}_2}(\mathbf{q})\Gamma^\alpha_{\mathbf{k}_1'\mathbf{k}_2'}(-\mathbf{q})\notag\\
    &(A^{\alpha0~\ast}_{\mathbf{k}_1'\mathbf{k}_2'}\!A^{\alpha0~\ast}_{\mathbf{k}_1\mathbf{k}_2}
    \delta_{\mathbf{k}_1'\!+\mathbf{k}_2'\!+\mathbf{q},\mathbf{k}_1\!+\!\mathbf{k}_2}\!\!\!-\!4A^{\alpha0~\ast}_{\mathbf{k}_2\mathbf{k}_2'}\!A^{\alpha0~\ast}_{\mathbf{k}_1,\mathbf{k}_1'}
    \delta_{\mathbf{k}_1',\mathbf{k}_2+\mathbf{q}}
    ),\label{eqn:nonlinearity}
\end{align}
where 
\begin{equation}
\Gamma^\alpha_{\mathbf{k}_1\mathbf{k}_2}(\mathbf{q})=A^{\alpha 0}_{\mathbf{k}_1+\mathbf{q},\mathbf{k}_2}+A^{\alpha 0}_{\mathbf{k}_1,\mathbf{k}_2+\mathbf{q}}-A^{\alpha0}_{\mathbf{k}_1\mathbf{k}_2}.    
\end{equation}
The full derivation of this result is given in Appendix \ref{app:nonlinearity}.

\section{Numerical implementation}\label{sec:numerical}

To calculate $g_{\mathrm{T}}$, we have to solve Eq.~\eqref{eqn:T-WannierQ} for some trion wavefunction. To do this, we expand coefficients using the Gaussian basis set \cite{Kidd:PRB93(2016),Ceferino:PRB101(2020)}
\begin{equation}
A^{\alpha 0}_{\mathbf{k}_1\mathbf{k}_2}=\sum_{\bm{n}}C^{\alpha}_{\bm{n}}\Upsilon^{\bm{n}}_{\mathbf{k}_1\mathbf{k}_2},\label{eqn:A_HO}
\end{equation}
where $\bm{n}=[n_{1x},n_{1y},n_{2x},n_{2y}]$ stands for the indices of a basis function,
\begin{equation}\label{eqn:expansion}
    \Upsilon^{\bm{n}}_{\mathbf{k}_1\mathbf{k}_2}=\varphi^{\lambda_1}_{\mathbf{n}_{1}}(\mathbf{k}_{1})\varphi^{\lambda_2}_{\mathbf{n}_{2}}(\mathbf{k}_{2}).
\end{equation}
We use basis functions of the form
\begin{equation}
\varphi^\lambda_{\mathbf{n}}(\mathbf{k})=\prod_{j=x,y}N_{n_j}\mathrm{e}^{-k_j^2\lambda^2/2}H_{n_j}(k_j\lambda)    
\end{equation}
where $H_{n_j}$ are Hermite polynomials \cite{Ceferino:PRB101(2020)}, and $N_n=\sqrt{\lambda/(\pi^{1/2}2^nn!)}$ is a normalization constant. In Eq.~\eqref{eqn:expansion}, the two-dimensional Hermite basis functions are labelled by the vector $\mathbf{n}$ with components $n_j$. 
Our basis set is characterized by two parameters being $\lambda_{1,2}$. In principle, they can be chosen arbitrary. However, to optimize the computational procedure we define their values using the procedure described shortly in the following. 
\begin{figure}
    \centering
    \includegraphics[width=3.3in]{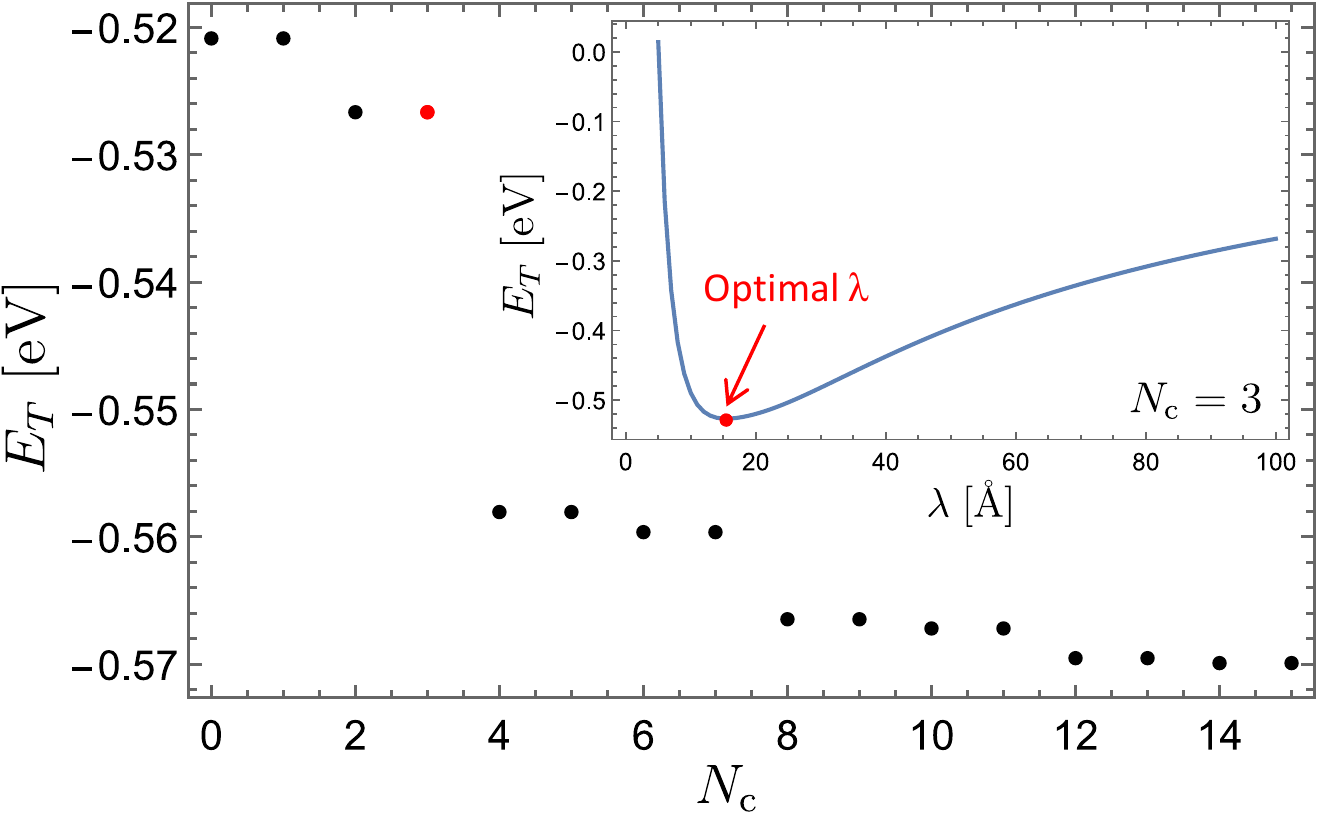}
    \caption{\textbf{Convergence of the trion energy} $E_{\mathrm{T}}$ (counted from the bandgap) with respect to the size of the basis set. In the chosen basis set all basis functions satisfy $n_{1x}+n_{1y}+n_{2x}+n_{2y}\leq N_\text{c}$ where $N_c$ is the cut-off parameter. In the main panel, $E_{\mathrm{T}}$ is shown for MoSe$_2$ parameters and $\lambda_1=\lambda_2=\lambda$ (red bullet) is the optimal length with $N_{\mathrm{c}}=3$ (see Fig.~\ref{fig:convergence}, inset).}
    \label{fig:convergence}
\end{figure}

With this orthogonal and complete basis we can rewrite Eq.~\eqref{eqn:T-WannierQ} as
\begin{align}\label{eqn:WannierHO}
\sum_{\bm{n}'}\Big[H_{0,\bm{n}\bm{n}'}+K_{\bm{n}\bm{n}'}\Big]C^\alpha_{\bm{n}'}=E_{\alpha }C^\alpha_{\bm{n}},
\end{align}
where
$H_{0,\bm{n}\bm{n}'}\!=\!\int_{\mathbf{k}} \Upsilon^{\bm{n}\ast}_{\mathbf{k}_1\mathbf{k}_2}[\varepsilon^{c}_{\mathbf{k}_1}\!+\varepsilon^{c}_{\mathbf{k}_2}\!-\varepsilon^{v}_{\mathbf{k}_1+\mathbf{k}_2}]\Upsilon^{\bm{n}}_{\mathbf{k}_1\mathbf{k}_2}
$
and 
$
K_{\bm{n}\bm{n}'}
=\frac{1}{L^2}\sum_{{\mathbf{q}_1\mathbf{q}_2}}\int_{\mathbf{k}}\Upsilon^{\bm{n}\ast}_{\mathbf{k}_1\mathbf{k}_2}\Xi_{\mathbf{q}_1\mathbf{q}_2}\Upsilon^{\bm{n}}_{\mathbf{k}_1+\mathbf{q},\mathbf{k}_2-\mathbf{q}}
$ with $\int_{\mathbf{k}}\equiv\int d\mathbf{k}_1d\mathbf{k}_2$. Solving Eq.~\eqref{eqn:WannierHO}, we obtain the trion wavefunction later used in Eq.~\eqref{eqn:A_HO}. 
To diagonalize Eq.~\eqref{eqn:WannierHO} numerically, we need to specify a total number of basis functions $N_{\mathrm{c}}$ and the parameters of the basis set, $\lambda_{1,2}$. In principle, various basis set choices with finite $\lambda_{1,2}$ is eventually converging to the unique solution, but the convergence rate differs depending on values of $\lambda_{1,2}$. To achieve better convergence, one needs to start with a small basis set and optimize the binding energy with respect to $\lambda_{1,2}$. Once this is done, the large basis set is constructed with this optimal $\lambda_{1,2}$ for the fast convergent calculation.

After obtaining the trion eigenfunction [Eq. \eqref{eqn:A_HO}], we can numerically calculate the corresponding Kerr nonlinearity coefficient as
\begin{align}
    g_{\mathrm{T}}=
    &\sum_{ \bm{m}\bm{m}'}\sum_{\bm{n}\bm{n}'}C^{\alpha}_{\bm{m}}C^{\alpha}_{\bm{m}'}C^{\alpha\ast}_{\bm{n}}C^{\alpha\ast}_{\bm{n}'}\Big[I^{\bm{m}\bm{m}'}_{\bm{n}\bm{n}'}-J^{\bm{m}\bm{m}'}_{\bm{n}\bm{n}'}\Big],
    \label{gT}
\end{align}
where $e-e$ exchange integral is
\begin{equation}
    I^{\bm{m}\bm{m}'}_{\bm{n}\bm{n}'}\!\!=\!-4\!\!\int_{\mathbf{k}}\!\int_{\mathbf{k}'}\!\!
    W(\mathbf{q})
    g^{\bm{m}}_{\mathbf{k}_1\mathbf{k}_2}(\mathbf{q})g^{\bm{m}'}_{\mathbf{k}_1'\mathbf{k}_2'}(-\mathbf{q})
    \Upsilon^{\bm{n}\ast}_{\mathbf{k}_1\mathbf{k}_1'}
    \Upsilon^{\bm{n}'\ast}_{\mathbf{k}_2\mathbf{k}_2'}\label{eqn:I}    
\end{equation}
with $\mathbf{q}=\mathbf{k}_2-\mathbf{k}_1'$, and $h-h$ exchange integral reads
\begin{align}
    J^{\bm{m}\bm{m}'}_{\bm{n}\bm{n}'}\!=\!
    \int_\mathbf{k}\!\int_{\mathbf{k}'}
    \!W(\mathbf{q})
    g^{\bm{m}}_{\mathbf{k}_1\mathbf{k}_2}(\mathbf{q})g^{\bm{m}'}_{\mathbf{k}_2'\mathbf{k}_1'}(-\mathbf{q})
    \Upsilon^{\bm{n}\ast}_{\mathbf{k}_1\mathbf{k}_2}\Upsilon^{\bm{n}'\ast}_{\mathbf{k}_1'\mathbf{k}_2'}\label{eqn:J}
\end{align}
with $\mathbf{q}=\mathbf{k}_1-\mathbf{k}_1'+\mathbf{k}_2-\mathbf{k}_2'$, and $g^{\bm{m}}_{\mathbf{k}_1\mathbf{k}_2}(\mathbf{q})=\Upsilon^{\bm{m}}_{\mathbf{k}_1+\mathbf{q},\mathbf{k}_2}+\Upsilon^{\bm{m}}_{\mathbf{k}_1,\mathbf{k}_2+\mathbf{q}}
-\Upsilon^{\bm{m}}_{\mathbf{k}_1\mathbf{k}_2}$. The momentum integration can be done analytically (see Appendix \ref{app:Iex}). Performing the summation in Eq.~\eqref{gT} can be computational demanding. However, we checked that most of the coefficients $C^\alpha_{\bm{m}}$ are small for the trion ground state, which substantially simplifies the procedure.

\section{Nonlinear response in 2D materials}

To analyze trion excitations in 2D materials, we first focus on one-trion properties. After this, we take into account the T-T interacting effects and investigate the related nonlinear response. 
\begin{table}[]
    \centering
    \caption{Model parameters for various TMD monolayers \cite{Kormanyos:2DMat2(2015),Danovich:PRB97(2018)} and their corresponding exciton ($E_X$) and trion ($E_b$) ground state binding energies (shown in [meV] units). $m_c$ and $m_v$ are the TMD conduction band and valence band masses ($m_0$ is the free electron mass). $r_\ast$ is the TMD screening length [\AA] of the Keldysh potential. The size of trion is characterized by the average distances [\AA] between electron-electron ($r_{\mathrm{ee}}$) and electron-hole ($r_{\mathrm{eh}}$) in a trion. The last column is the nonlinearity due to T-T interaction, $g_{\mathrm{T}}$ [$\mu\mathrm{eV}\mu\mathrm{m}^2$]. 
    }
    \label{tbl:TMD}
    \begin{ruledtabular}
    \begin{tabular}{c|cccccccc}
     & $m_c/m_0$&$m_v/m_0$&$r_\ast$ &$E_X$ & $E_b$&$r_{\mathrm{ee}}$&$r_{\mathrm{eh}}$&$g_{\mathrm{T}}$\\
    \hline
       MoSe$_2$ & 0.38 & 0.44 & 40 & 539 & 31 & 27 & 17&$-2.8$\\
       MoS$_2$  & 0.47 & 0.54 & 35 & 628 & 35 & 22 & 14& $-2.3$\\
       WSe$_2$  & 0.29 & 0.34 & 40 & 549 & 33 & 29 & 19&$-3.2$\\
       WS$_2$   & 0.27 & 0.36 & 45 & 458 & 27 & 31 & 21& $-3.7$ \\
    \end{tabular}
    \end{ruledtabular}
\end{table}

\subsection{One-trion properties}
In an optical microcavity only low-energy trion excitations ($\mathbf{Q}\approx0$) couple to light, forming a trion-polariton. For small $\mathbf{Q}$ and trion ground state only the electronic modes near the band edges are relevant~\cite{Zhumagulov:NPJ2022}. Therefore, we can use an effective mass approximation for the dispersion of the conduction and the valence bands, $\varepsilon_c(\mathbf{k})=k^2/2m_c$ and $\varepsilon_v(\mathbf{k})=-k^2/2m_v$. The interaction potential has to account for the effects of 2D screening~\cite{Keldysh:JETP29-1979,Rytova:MUPhys3-1967} and can be written as
\begin{equation}\label{eqn:W}
    W(\mathbf{q})=\frac{2\pi e^2}{ q(1+r_\ast q)},
\end{equation}
where $r_\ast$ is a screening parameter.

We analyze the one-trion state for monolayer transition metal dichalcogenides (TMD) by solving the Wannier equation in Eq.~\eqref{eqn:T-WannierQ}. The trion energy $E_{\mathrm{T}}$ converges to a desirable level of accuracy with the cutoff parameter $N_{\mathrm{c}}=12$ (1820 basis functions), as demonstrated in Fig.~\ref{fig:convergence}. We set $\lambda_1=\lambda_2=\lambda$ for simplicity and the optimal $\lambda$ is obtained by minimizing $E_{\mathrm{T}}$ with for small value of $N_\text{c}=3$ (see the inset in Fig.~\ref{fig:convergence}).

In Table \ref{tbl:TMD}, we summarize parameters of TMD materials and results from our calculations of the one-trion properties. We characterize trions by the binding energy ($E_b$), average electron-electron distance ($r_{\mathrm{ee}}$) and average electron-hole distance ($r_{\mathrm{eh}}$). The average distance is calculated by using $r_{ij}=\langle|\mathbf{r}_i-\mathbf{r}_j|\rangle_{\mathrm{T}}$ where $\langle\dots\rangle_{\mathrm{T}}$ stands for the average over the trion wavefunction in real space. We note that different TMDs have similar exciton $E_X$ and trion $E_b$ binding energies, as well as trion sizes ($r_{\mathrm{ee}}$ and $r_{\mathrm{eh}}$). 
\begin{figure}
    \centering
    \includegraphics[width=3.3in]{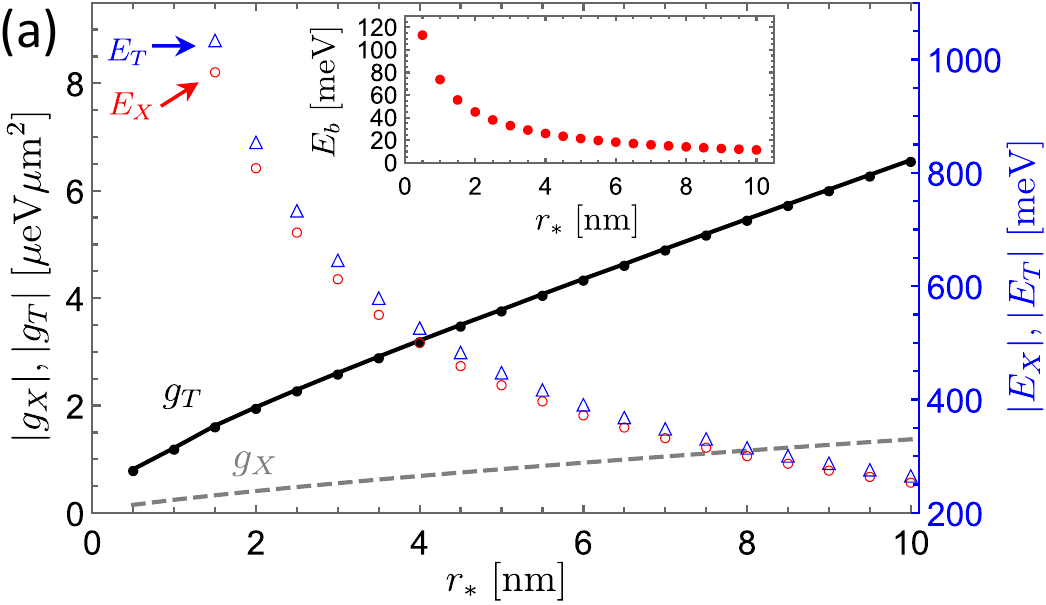}
    \includegraphics[width=3.3in]{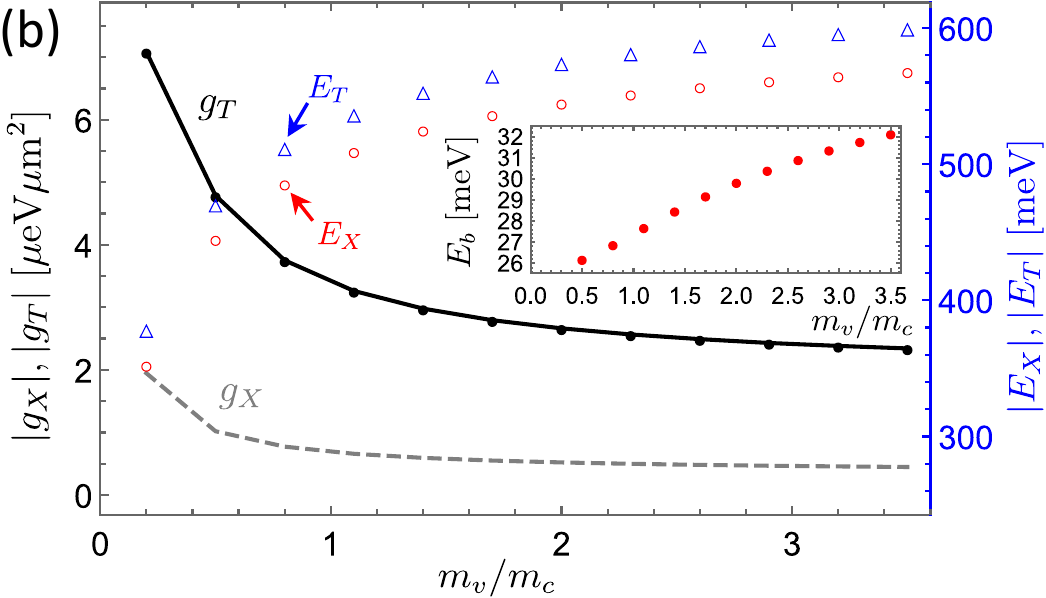}
    \caption{\textbf{Nonlinear interaction coefficients.} (a) Nonlinearity due T-T interactions (black solid curve, $g_{\mathrm{T}}<0$) and X-X interactions (gray dashed curve, $g_X>0$) plotted as a function of screening parameter $r_{\ast}$. Red circles and blue triangles correspond to exciton ($E_X$) and trion ($E_{\mathrm{T}}$) energies, where these energies are measured from the band gap. In calculations we set $m_v=m_c=0.32 m_0$. The trion binding energy $E_b=|E_{\mathrm{T}}-E_X|$ is plotted in the inset. (b) Nonlinear coefficients plotted as a function of the hole-to-electron mass ratio, $m_v/m_e$. Notation is the same and we use $r_\ast=4$~nm.}
    \label{fig:gT}
\end{figure}

\subsection{Nonlinearity due to T-T interactions}\label{sec:nonlinearity}

After calculating the single-trion properties of trions, we proceed with evaluating the matrix elements of the direct and exchange trion-trion scattering and estimating the corresponding Kerr nonlinearity coefficient. The values of $g_{\mathrm{T}}$ for considered materials are given in the last column of Table~\ref{tbl:TMD}. We present the dependence of $g_{\mathrm{T}}$ on screening length $r^\ast$ and the hole-to-electron mass ratio $m_v/m_c$ in Fig.~\ref{fig:gT}. The important result of our calculations is the prediction of a negative sign for the nonlinear interaction coefficient for trions. Consequently, this leads to the attraction between trion-polaritons in TMD monolayers, as shown in our accompanying letter \cite{Song2022}. The attractive T-T interaction, which induces nonlinear redshift for ground state trion-polaritons, goes in striking contrast to ground state X-X scattering that leads to the blueshift. This implies possible qualitative changes for the nonlinear optical effects in polaritonic systems, arising from focusing on nonlinearity. As a possible origin of attraction, we consider the non-monotonic spatial dependence in the trion wavefunction (unlike 1s excitons), similarly to the case of excited X-X interaction \cite{Shahnazaryan:PRB102(2020),Shahnazaryan2016}. We also stress that the absolute value of the Kerr nonlinearity coefficient is substantially larger for trion-polaritons than for the exciton-polaritons. This is due to their bigger size favorable for the electron-electron and hole-hole exchange processes. Our results thus confirm trion-polaritons as a favorable system for increasing the nonlinear response.

\section{Conclusion}\label{sec:conclusion}

To conclude, we developed a quantitative microscopic theory for the nonlinear optical response of 2D semiconductors with free electrons, accounting for the composite nature of photo-created excitations correlated with the Fermi sea. The response of such system can be described in terms of trion-based excitations, which dominate in the regime where binding energy is bigger than the Fermi energy.
We described the leading microscopic processes constituting the trion-trion interaction and presented a derivation of the nonlinear T-T interaction coefficient. We estimated numerically the nonlinear coefficient by using the Gaussian basis function method for describing the trion wavefunction. The results were analyzed for various TMD materials and studied as a function of screening parameters, identifying that the increase in screening leads to the linear growth of the nonlinear coefficient. Importantly, our calculations predict trion-trion attraction (see accompanying Ref.~\cite{Song2022}). This results in a redshift of the trion line in the spectrum, in contrast to the typical blueshift for excitons. The developed theoretical approach shows the recipes required for studying correlations with composite multi-body quasiparticles and opens a route to studies of optical nonlinearity in doped 2D materials.

\begin{acknowledgements}
The authors acknowledge the support from UK EPSRC New Investigator Award under the Agreement No. EP/V00171X/1. S.\,C. and O.\,K. are supported by the NATO Science for Peace and Security project NATO.SPS.MYP.G5860. I.\,A.\,S. acknowledges support from IRF (project ``Hybrid polaritonics'') and RFBR, project No. 21-52-12038.
\end{acknowledgements}

\appendix

\section{Commutation relations for $\mathcal{T}^\dagger_{\alpha\mathbf{Q}}$ and $\mathcal{T}_{\alpha\mathbf{Q}}$
    }\label{app:commutator}

In this section, we summarize the commutation relations for the trion field operators. To prove these relations, 
the completeness relations for a trion wavefunction
\begin{align}
\sum_{\alpha'}A^{\alpha' \mathbf{Q}}_{\mathbf{k}'_1\mathbf{k}'_2}A^{\alpha'\mathbf{Q},\ast}_{\mathbf{k}_1\mathbf{k}_2}&=\delta_{\mathbf{k}_1'\mathbf{k}_1}\delta_{\mathbf{k}_2'\mathbf{k}_2},\\
\sum_{\mathbf{k}_1\mathbf{k}_2}A^{\alpha \mathbf{Q}}_{\mathbf{k}_1\mathbf{k}_2}A^{\alpha'\mathbf{Q},\ast}_{\mathbf{k}_1\mathbf{k}_2}&=\delta_{\alpha'\alpha}
\end{align} 
are particularly useful. To derive these completeness relations, one may use the fact that Eq.~\eqref{eqn:T-WannierQ}
is a linear equation with real eigenvalues. With these completeness, we can represent a non-interaction three-particle states as
\begin{equation}\label{eqn:b2f}
    \sum_{\alpha}A^{\alpha\mathbf{Q},\ast}_{\mathbf{k}_1\mathbf{k}_2}\mathcal{T}^\dagger_{\alpha\mathbf{Q}}=\frac{1}{\sqrt{2!}}a^\dagger_{\mathbf{k}_1}a^\dagger_{\mathbf{k}_2}b_{\mathbf{k}_1+\mathbf{k}_2-\mathbf{Q}}.
\end{equation}
The relation above leads to the following important commutation rules:
\begin{align}
    [a^\dagger_{\mathbf{k}+ \mathbf{q}}a_{\mathbf{k}},\mathcal{T}^\dagger_{\alpha\mathbf{Q}}]
    \!=\!&
     \sum_{\bar{\alpha}\mathbf{k}_1\mathbf{k}_2}\!\!2A^{\alpha\mathbf{Q}}_{\mathbf{k}_1,\mathbf{k}_2}A^{\bar{\alpha}\mathbf{Q}+\mathbf{q},\ast}_{\mathbf{k}_1+\mathbf{q},\mathbf{k}_2}\mathcal{T}^\dagger_{\bar{\alpha}\mathbf{Q}+\mathbf{q}}\delta_{\mathbf{k},\mathbf{k}_1}\label{eqn:[aa,T]}\\
    [b^\dagger_{\mathbf{k}+\mathbf{q}}b_{\mathbf{k}},\mathcal{T}^\dagger_{\alpha\mathbf{Q}}]
    \!=\!&
    -\!\!\!\sum_{\bar{\alpha}\mathbf{k}_1\mathbf{k}_2}\!\!\!A^{\alpha \mathbf{Q}}_{\mathbf{k}_1\mathbf{k}_2}A^{\bar{\alpha}\mathbf{Q}+\mathbf{q},\ast}_{\mathbf{k}_1\mathbf{k}_2}\mathcal{T}^\dagger_{\bar{\alpha}\mathbf{Q}+\mathbf{q}}\!\delta_{\mathbf{k},\mathbf{k}_v\!-\mathbf{q}}\label{eqn:[bb,T]}
\end{align}
where we remind that the valence band momentum is $\mathbf{k}_v=\mathbf{k}_1+\mathbf{k}_2-\mathbf{Q}$. The above two commutation relations lead to
\begin{equation}
     [\rho_{\mathbf{q}},\mathcal{T}^\dagger_{\alpha\mathbf{Q}}]=\sum_{\alpha'}
     F^{\alpha\alpha'}_{\mathbf{Q},\mathbf{Q}+\mathbf{q}}
    \mathcal{T}^\dagger_{\alpha',\mathbf{Q}+\mathbf{q}},
\end{equation}
with $F^{\alpha\alpha'}_{\mathbf{Q},\mathbf{Q}+\mathbf{q}}\!=\!\displaystyle\sum_{\mathbf{k}_1\mathbf{k}_2}[A^{\alpha\mathbf{Q}}_{\mathbf{k}_1-\mathbf{q},\mathbf{k}_2}\!+\!A^{\alpha\mathbf{Q}}_{\mathbf{k}_1,\mathbf{k}_2-\mathbf{q}}\!-\!A^{\alpha \mathbf{Q}}_{\mathbf{k}_1\mathbf{k}_2}]A^{\alpha',\mathbf{Q}+\mathbf{q},\ast}_{\mathbf{k}_1\mathbf{k}_2}$ being the trion wavefunction overlapping factor. This yields Eq. \eqref{eqn:[H,T]}.

To calculate the exchange processes, the important commutation rule is
\begin{equation}\label{eqn:[T,T]}
    \{\mathcal{T}_{\alpha'\mathbf{Q}'},\mathcal{T}^\dagger_{\alpha\mathbf{Q}}\}=\delta_{\alpha'\alpha}\delta_{\mathbf{Q},\mathbf{Q}}+\mathcal{D}^{\alpha'\alpha}_{\mathbf{Q}'\mathbf{Q}}
\end{equation}
where the $\mathcal{D}^{\alpha'\alpha}_{\mathbf{Q}'\mathbf{Q}}$ arises from the composite nature of trionic field and reads
\begin{align*}
&\mathcal{D}^{\alpha'\alpha}_{\mathbf{Q}'\mathbf{Q}}=-
    \sum_{\mathbf{k}_1'\mathbf{k}_2'}\sum_{\mathbf{k}_1\mathbf{k}_2}A^{\alpha'\mathbf{Q}',\ast}_{\mathbf{k}_1'\mathbf{k}_2'}A^{\alpha \mathbf{Q}}_{\mathbf{k}_1\mathbf{k}_2}\Big[ \delta_{\mathbf{k}_1'\mathbf{k}_1}\big(\delta_{\mathbf{k}_2'\mathbf{k}_2}b_{\mathbf{k}_v}b^\dagger_{\mathbf{k}_v'}\notag\\
    &+2a^\dagger_{\mathbf{k}_2}a_{\mathbf{k}_2'}(\delta_{\mathbf{k}_v',\mathbf{k}_v}-
    b_{\mathbf{k}_v}b^\dagger_{\mathbf{k}_v'})\big)
    -\frac{1}{2}a^\dagger_{\mathbf{k}_1}a^\dagger_{\mathbf{k}_2}a_{\mathbf{k}_2'}a_{\mathbf{k}_1'}\delta_{\mathbf{k}_v',\mathbf{k}_v}\Big]
\end{align*}
where the valence band momenta in the above are $\mathbf{k}_v=\mathbf{k}_1+\mathbf{k}_2-\mathbf{Q}$ and $\mathbf{k}_v'=\mathbf{k}_1'+\mathbf{k}_2'-\mathbf{Q}'$. To evaluate the commutator in Eq.~\eqref{eqn:[T,T]}, the following identities $[A,BC]=\{A,B\}C-B\{A,C\}$ and $[BC,A]=B\{A,C\}-\{A,B\}C$ are useful. 

Furthermore, using Eqs.~\eqref{eqn:[aa,T]} and \eqref{eqn:[bb,T]}, we obtain the other commutation relation 
\begin{align}
    [\mathcal{D}^{\alpha'\alpha}_{\mathbf{Q}'\mathbf{Q}},T^\dagger_{\beta\mathbf{P}}]=\sum_{\bar{\alpha} \bar{\mathbf{Q}}}&\mathcal{T}^\dagger_{\bar{\alpha}\bar{\mathbf{Q}}}\Big[\Big(\Lambda^{\alpha\beta,\alpha'\bar{\alpha}}_{\mathbf{Q}\mathbf{P},\mathbf{Q}'\bar{\mathbf{Q}}}-\Lambda^{\alpha\beta,\bar{\alpha}\alpha'}_{\mathbf{Q}\mathbf{P},\bar{\mathbf{Q}}\mathbf{Q}'}\Big)\notag\\
    &\times\delta_{\bar{\mathbf{Q}},\mathbf{Q}-\mathbf{Q}'+\mathbf{P}}+\mathcal{P}^{\alpha\beta,\alpha'\bar{\alpha}}_{\mathbf{Q}\mathbf{P},\mathbf{Q}'\bar{\mathbf{Q}}}\Big]
\end{align}
with the Pauli's scattering $\Lambda^{\alpha\beta,\alpha'\bar{\alpha}}_{\mathbf{Q}\mathbf{P},\mathbf{Q}'\bar{\mathbf{Q}}}$ [Eq.~\eqref{eqn:Lambda}], and
\begin{align*}
    &\mathcal{P}^{\alpha\beta,\alpha'\bar{\alpha}}_{\mathbf{Q}\mathbf{P},\mathbf{Q}'\bar{\mathbf{Q}}}=\sum_{\mathbf{k}_1'\mathbf{k}_2'}\sum_{\mathbf{k}_1\mathbf{k}_2}\sum_{\mathbf{p}_1\mathbf{p}_2}\sum_{\bar{\mathbf{p}}_1\bar{\mathbf{p}}_2}A^{\alpha'\mathbf{Q}',\ast}_{\mathbf{k}_1'\mathbf{k}_2'}A^{\alpha \mathbf{Q}}_{\mathbf{k}_1\mathbf{k}_2}A^{\beta\mathbf{P}}_{\mathbf{p}_1\mathbf{p}_2}A^{\bar{\alpha}\bar{\mathbf{Q}}\ast}_{\bar{\mathbf{p}}_1\bar{\mathbf{p}}_2}\notag\\
    &\Big[\delta_{\mathbf{k}_1'\mathbf{k}_1}\delta_{\mathbf{p}_2\bar{\mathbf{p}}_2}\Big(4\delta_{\mathbf{k}_2',\mathbf{p}_1}\delta_{\bar{\mathbf{p}}_1,\mathbf{p}_1+\mathbf{k}_2-\mathbf{k}_2'}
    \delta_{\bar{\mathbf{Q}},\mathbf{P}+\mathbf{k}_2-\mathbf{k}_2'}
    b_{\mathbf{k}_v}b^\dagger_{\mathbf{k}_v'}
    \notag\\
    &+
    2\delta_{\mathbf{p}_1\bar{\mathbf{p}}_1}
    \delta_{\mathbf{k}_v,\bar{\mathbf{p}}_v}\delta_{\bar{\mathbf{Q}},\mathbf{P}+\mathbf{k}_v'-\mathbf{k}_v}a^\dagger_{\mathbf{k}_2}a_{\mathbf{k}_2'}
    \Big)\notag\\
    &
    -2\delta_{\bar{\mathbf{p}}_1\mathbf{k}_1}\delta_{\bar{\mathbf{p}}_2\mathbf{k}_2}\delta_{\bar{\mathbf{Q}},\mathbf{k}_1+\mathbf{k}_2-\mathbf{p}_v}\delta_{\mathbf{p}_1\mathbf{k}_1'}\delta_{\mathbf{k}_v',\mathbf{k}_v}a^\dagger_{\mathbf{p}_2}a_{\mathbf{k}_2'}\Big],
\end{align*}
where the valence band momenta are $\mathbf{p}_v=\mathbf{p}_1+\mathbf{p}_2-\mathbf{P}$ and $\bar{\mathbf{p}}_v=\bar{\mathbf{p}}_1+\bar{\mathbf{p}}_2-\bar{\mathbf{Q}}$.


\section{Derivation of nonlinearity interaction coefficients}\label{app:nonlinearity}

To evaluate $\Delta^{(N)}_{\alpha\mathbf{K}}$, 
we use the commutation relation
\begin{equation}\label{eqn:[B^N,A]}
    [B^N,A]=\sum_{i=1}^{N}\binom{N}{i}\mathcal{L}_B^i(A)B^{N-i} ,
\end{equation}
where we have used the notation $\mathcal{L}_B(\bullet)=[B,\bullet]$. This identity can be proved by induction with the use of hockey-stick identity, $\sum_{i=k}^N\binom{i}{k}=\binom{N+1}{k+1}$ with $\binom{j}{i}=0$ if $j<i$. Using Eq.~\eqref{eqn:[H,B]} and Eq.~\eqref{eqn:[B^N,A]}, we obtain
\begin{widetext}
\begin{align}
    \langle N|\mathcal{H}_e|N\rangle= NE_{\alpha\mathbf{K}}+\frac{1}{L^2N_{\mathrm{F}}}\frac{F_{N-2}}{2F_N}\sum_{\mathbf{q}}W(\mathbf{q})\sum_{i=1}^N\binom{N}{i} \frac{\langle\varnothing|\mathcal{L}^i_{\mathcal{B}_{\alpha\mathbf{K}}}[\mathcal{C}^\dagger_{\alpha\mathbf{K}}(-\mathbf{q})\mathcal{C}^\dagger_{\alpha\mathbf{K}}(\mathbf{q})]}{\sqrt{(N-2)!F_{N-2}}}\mathcal{B}^{N-i}_{\alpha\mathbf{K}}|N-2\rangle .
\end{align}
Keeping only $i=1,2$ in Eq.~\eqref{eqn:[B^N,A]}, since $i\geq3$ corresponds to the exchange processes with more than 2 trions. This leads to Eq.~\eqref{eqn:<N|H|N>}. To compute the expectation value in Eq.~\eqref{eqn:<N|H|N>}, we rewrite it as
\begin{align}
    \langle\varnothing|
    \mathcal{B}^2_{\alpha\mathbf{K}}\mathcal{C}^\dagger_{\alpha,\mathbf{K}}(-\mathbf{q})\mathcal{C}^\dagger_{\alpha,\mathbf{K}}(\mathbf{q})
    \mathcal{B}_{\alpha\mathbf{K}}^{N-2}| N-2\rangle
    \approx&\frac{1}{N_{\mathrm{F}}}\sum_{\mathbf{p}_1\mathbf{p}_2\mathbf{p}_3\mathbf{p}_4}
    \sum_{\alpha'\beta'}\Theta_{k_{\mathrm{F}}}(|\mathbf{p}_1|)\Theta_{k_{\mathrm{F}}}(|\mathbf{p}_2|)G^{\alpha\alpha'}_{\mathbf{K}}(\mathbf{p}_3,-\mathbf{q})G^{\alpha\beta'}_{\mathbf{K}}(\mathbf{p}_4,\mathbf{q})\notag\\
    &\langle\varnothing|a^\dagger_{\mathbf{p}_1}a^\dagger_{\mathbf{p}_2}
    \mathcal{T}_{\alpha\tilde{\mathbf{p}}_1}\mathcal{T}_{\alpha\tilde{\mathbf{p}}_2}\mathcal{T}^\dagger_{\alpha',\tilde{\mathbf{p}}_3-\mathbf{q}}\mathcal{T}^\dagger_{\beta',\tilde{\mathbf{p}}_4+\mathbf{q}}a_{\mathbf{p}_3}a_{\mathbf{p}_4}
    \mathcal{B}_{\alpha\mathbf{K}}^{N-2}| N-2\rangle ,
\end{align}
where we have let $\{\mathcal{T}^\dagger_{\beta',\tilde{\mathbf{p}}_4+\mathbf{q}},a_{\mathbf{p}_3}\}\approx0$ in the small $k_{\mathrm{F}}$ limit.

To proceed further, we then perform operator ordering to use relations $\mathcal{T}_{\alpha\mathbf{Q}}|\varnothing\rangle=0$, $\{\mathcal{T}_{\alpha'\mathbf{Q}'},\mathcal{T}^\dagger_{\alpha\mathbf{Q}}\}$ in Eq.~\eqref{eqn:[T,T]}, and $[\mathcal{D}^{\alpha'\alpha}_{\mathbf{Q}'\mathbf{Q}},T^\dagger_{\beta\mathbf{P}}]$ in Eq.~\eqref{eqn:[D,T]}. We let  $\mathcal{D}^{\alpha'\alpha}_{\mathbf{Q}'\mathbf{Q}}|\varnothing\rangle=0$ and $\langle \varnothing|a^\dagger_{\mathbf{p}_1}a^\dagger_{\mathbf{p}_2} a_{\mathbf{p}_3}a_{\mathbf{p}_4}\mathcal{B}_{\alpha\mathbf{K}}^{N-2}|N-2\rangle/\sqrt{(N-2)!F_{N-2}}\approx\delta_{\mathbf{p}_1\mathbf{p}_4}\delta_{\mathbf{p}_2\mathbf{p}_3}-\delta_{\mathbf{p}_1\mathbf{p}_3}\delta_{\mathbf{p}_2\mathbf{p}_4}$ in the $k_{\mathrm{F}}\sim0$ limit. Therefore, we obtain Eq.~\eqref{eqn:gT} presented in the main text.

Evaluating $g_{\mathrm{T}}$ without knowing an analytical expression for trion wavefunction is a challenging task, and we proceed by doing it numerically for leading contributions in Eq.~\eqref{eqn:gT}. To do this, we assume that the trion wavefunction is a slow-varying function of its total momentum. Therefore, we can rewrite exchange energies as
\begin{align}
    \bar{V}_{\alpha\alpha;\alpha\alpha}^{\mathbf{p}_1\mathbf{p}_2,\mathbf{p}_1-\mathbf{q},\mathbf{p}_2+\mathbf{q}}
    \approx&\sum_{\mathbf{q}'}
    \sum_{\mathbf{k}_1\mathbf{k}_2}\sum_{\mathbf{k}_1'\mathbf{k}_2'}W(\mathbf{q}')
    \Big(A^{\alpha0~\ast}_{\mathbf{k}_1'\mathbf{k}_2'}\!A^{\alpha0~\ast}_{\mathbf{k}_1\mathbf{k}_2}
    \delta_{\mathbf{k}_1'+\mathbf{k}_2'+\mathbf{q}',\mathbf{k}_1+\mathbf{k}_2-\mathbf{q}}-4A^{\alpha0~\ast}_{\mathbf{k}_2\mathbf{k}_2'}\!A^{\alpha0~\ast}_{\mathbf{k}_1,\mathbf{k}_1'}
    \delta_{\mathbf{k}_1'+\mathbf{q}',\mathbf{k}_2-\mathbf{q}}
    \Big)
    \notag\\
    &
    \Big(\Gamma^\alpha_{\mathbf{k}_1\mathbf{k}_2}(\mathbf{q}')-\Theta_{k_{\mathrm{F}}}(|\mathbf{p}_1+\mathbf{q}'|)A^{\alpha0}_{\mathbf{k}_1\mathbf{k}_2}\Big)\Big(\Gamma^\alpha_{\mathbf{k}_1'\mathbf{k}_2'}(-\mathbf{q}')-\Theta_{k_{\mathrm{F}}}(|\mathbf{p}_2-\mathbf{q}'|)A^{\alpha0}_{\mathbf{k}_1'\mathbf{k}_2'}\Big)
\end{align}
where $\Gamma^\alpha_{\mathbf{k}_1\mathbf{k}_2}(\mathbf{q})=A^{\alpha 0}_{\mathbf{k}_1+\mathbf{q},\mathbf{k}_2}+A^{\alpha 0}_{\mathbf{k}_1,\mathbf{k}_2+\mathbf{q}}-A^{\alpha0}_{\mathbf{k}_1\mathbf{k}_2}$. 
This gives 
\begin{align}
    g_{\mathrm{T}}\approx&\frac{1}{N_{\mathrm{F}}^2}{\sum_{\mathbf{p}_1\mathbf{p}_2}}'\Big[
    \bar{V}_{\alpha\alpha;\alpha\alpha}^{\mathbf{p}_1\mathbf{p}_2,\mathbf{p}_1\mathbf{p}_2}
    +
    \bar{U}_{\alpha\alpha;\alpha\alpha}^{\mathbf{p}_1\mathbf{p}_2,\mathbf{p}_2\mathbf{p}_1}
    -
    \bar{V}_{\alpha\alpha;\alpha\alpha}^{\mathbf{p}_1\mathbf{p}_2,\mathbf{p}_2\mathbf{p}_1} ,
    \Big]\label{eqn:gT_VV'}
\end{align}
where we note that the direct interaction $\bar{U}_{\alpha\alpha,\alpha\alpha}^{\mathbf{p}_1\mathbf{p}_2,\mathbf{p}_1\mathbf{p}_2}=0$.
We further rewrite
\begin{align}
    \bar{V}_{\alpha\alpha;\alpha\alpha}^{\mathbf{p}_1\mathbf{p}_2,\mathbf{p}_1,\mathbf{p}_2}
    =&\sum_{\mathbf{q}}
    \sum_{\mathbf{k}_1\mathbf{k}_2}\sum_{\mathbf{k}_1'\mathbf{k}_2'}W(\mathbf{q})
    \Big(A^{\alpha0~\ast}_{\mathbf{k}_1'\mathbf{k}_2'}\!A^{\alpha0~\ast}_{\mathbf{k}_1\mathbf{k}_2}
    \delta_{\mathbf{k}_1'+\mathbf{k}_2'+\mathbf{q},\mathbf{k}_1+\mathbf{k}_2}-4A^{\alpha0~\ast}_{\mathbf{k}_2\mathbf{k}_2'}\!A^{\alpha0~\ast}_{\mathbf{k}_1,\mathbf{k}_1'}
    \delta_{\mathbf{k}_1',\mathbf{k}_2+\mathbf{q}}
    \Big)\Big[
    \Gamma^\alpha_{\mathbf{k}_1\mathbf{k}_2}(\mathbf{q})\Gamma^\alpha_{\mathbf{k}_1'\mathbf{k}_2'}(-\mathbf{q})
    \notag\\
    &
    -\Gamma^\alpha_{\mathbf{k}_1\mathbf{k}_2}(\mathbf{q})\Theta_{k_{\mathrm{F}}}(|\mathbf{p}_2-\mathbf{q}|)-\Gamma^\alpha_{\mathbf{k}_1'\mathbf{k}_2'}(-\mathbf{q})\Theta_{k_{\mathrm{F}}}(|\mathbf{p}_1+\mathbf{q}|)+\Theta_{k_{\mathrm{F}}}(|\mathbf{p}_2-\mathbf{q}|)\Theta_{k_{\mathrm{F}}}(|\mathbf{p}_1+\mathbf{q}|)\Big] .
    \label{eqn:V0}
\end{align}
We can see that the last two terms in Eq.~\eqref{eqn:gT_VV'} and the last line in Eq.~\eqref{eqn:V0} are $\mathbf{p}_{1,2}$-dependent. Hence, these terms are suppressed by $1/N_{\mathrm{F}}$-factor as compared to the first line in Eq.~\eqref{eqn:gT_VV'}. Therefore, we ignore these sub-leading terms for simplicity, since these terms are high-dimensional integrals which cannot be rewritten and evaluated analytically. Keeping only the first line in Eq.~\eqref{eqn:V0}, we obtain the nonlinearity in the main text Eq.~\eqref{eqn:nonlinearity}.


\section{Exchange integral $I$ and $J$}\label{app:Iex}

To integrate out momentum $k$, we make the change of variables as $\begin{bmatrix}\mathbf{k}_1'\\\mathbf{k}_2\end{bmatrix}=\begin{bmatrix}\mathbf{p}+\frac{1}{2}\mathbf{q}\\\mathbf{p}-\frac{1}{2}\mathbf{q}\end{bmatrix}$. This gives 
\begin{align}
    I^{\bm{m}\bm{m}'}_{\bm{n}\bm{n}'}=&
    -4\int d\mathbf{p}d\mathbf{q} d\mathbf{k}_1 d\mathbf{k}_2'
    W(\mathbf{q})
    g^{\bm{m}}_{\mathbf{k}_1,\mathbf{p}-\frac{1}{2}\mathbf{q}}(\mathbf{q})g^{\bm{m}'}_{\mathbf{p}+\frac{1}{2}\mathbf{q},\mathbf{k}_2'}(-\mathbf{q})
    \Upsilon^{\bm{n}\ast}_{\mathbf{k}_1,\mathbf{p}+\frac{1}{2}\mathbf{q}}\Upsilon^{\bm{n}'\ast}_{\mathbf{p}-\frac{1}{2}\mathbf{q},\mathbf{k}_2'}\notag\\
    =&-4\int d\mathbf{q} W(\mathbf{q})\Big[(f^{\lambda}_{\mathbf{m}_2',\mathbf{n}_2'}(-\mathbf{q})-\delta_{\mathbf{m}_2'\mathbf{n}_2'})(f^{\lambda}_{\mathbf{m}_1\mathbf{n}_1}(\mathbf{q})-\delta_{\mathbf{m}_1\mathbf{n}_1})h_{\mathbf{m}_1'\mathbf{n}_1',\mathbf{m}_2\mathbf{n}_2}(\mathbf{q})+\delta_{\mathbf{m}_2'\mathbf{n}_2'}\delta_{\mathbf{m}_1\mathbf{n}_1}h_{\mathbf{m}_2\mathbf{n}_1',\mathbf{m}_1'\mathbf{n}_2}(\mathbf{q})\notag\\
    &
    +
    \delta_{\mathbf{m}_2'\mathbf{n}_2'}(f^{\lambda}_{\mathbf{m}_1\mathbf{n}_1}(\mathbf{q})-\delta_{\mathbf{m}_1\mathbf{n}_1})h'_{\mathbf{m}_1'\mathbf{n}_1',\mathbf{m}_2\mathbf{n}_2}(\mathbf{q})
    +
    (f^{\lambda}_{\mathbf{m}_2',\mathbf{n}_2'}(-\mathbf{q})-\delta_{\mathbf{m}_2'\mathbf{n}_2'})\delta_{\mathbf{m}_1\mathbf{n}_1}h'_{\mathbf{m}_1'\mathbf{n}_2,\mathbf{m}_2\mathbf{n}_1'}(-\mathbf{q})
    \Big] ,
\end{align}
where $\delta_{\mathbf{n}\mathbf{m}}=\delta_{n_xm_x}\delta_{n_ym_y}$,
\begin{equation}
    f^\lambda_{\mathbf{n}\mathbf{n}'}(\mathbf{q})=\prod_{i=x,y}\sum_{s_i=0}^{\text{min}[n_i,n_i']}\frac{(-1)^{n_i-s_i}2^{s_i}s_i!}{\sqrt{2^{n_i+n_i'}n_i!n_i'!}}\binom{n_i}{s_i}\binom{n_i'}{s_i}(q_i\lambda)^{n_i+n_i'-2s_i}\mathrm{e}^{-\frac{1}{4}q_i^2 \lambda^2},
\end{equation}
and
\begin{align}
    h_{\mathbf{m}_1'\mathbf{n}_1',\mathbf{m}_2\mathbf{n}_2}(\mathbf{q})=&\int d\mathbf{p}\varphi^{\lambda}_{\mathbf{m}_1'}(\mathbf{p}+\tfrac{1}{2}\mathbf{q})\varphi^{\lambda\ast}_{\mathbf{n}_1'}(\mathbf{p}-\tfrac{1}{2}\mathbf{q})\varphi^{\lambda}_{\mathbf{m}_2}(\mathbf{p}-\tfrac{1}{2}\mathbf{q})\varphi^{\lambda\ast}_{\mathbf{n}_2}(\mathbf{p}+\tfrac{1}{2}\mathbf{q}),\\
    h'_{\mathbf{m}_1'\mathbf{n}_1',\mathbf{m}_2\mathbf{n}_2}(\mathbf{q})=&\int d\mathbf{p}\varphi^{\lambda}_{\mathbf{m}_1'}(\mathbf{p}-\tfrac{1}{2}\mathbf{q})\varphi^{\lambda\ast}_{\mathbf{n}_1'}(\mathbf{p}-\tfrac{1}{2}\mathbf{q})\varphi^{\lambda}_{\mathbf{m}_2}(\mathbf{p}-\tfrac{1}{2}\mathbf{q})\varphi^{\lambda\ast}_{\mathbf{n}_2}(\mathbf{p}+\tfrac{1}{2}\mathbf{q}).
\end{align}
Here, for simplicity, we have chosen $\lambda_1=\lambda_2=\lambda$. It is a good basis set if the two electrons are identical. In this case,
we can further integrate out the momentum by using
\begin{align}
     H_{m}(x+y)H_{n}(x-y)
    =&\sum_{s=0}^m\sum_{r=0}^{n}\frac{(-1)^{r}}{\sqrt{2}^{m+n}}\binom{m}{s}\binom{n}{r}H_{m+n-s-r}(\sqrt{2}x)H_{r+s}(\sqrt{2}y).
\end{align}
Integrating out $\mathbf{p}$, we get
\begin{align}
    h_{\mathbf{m}_1'\mathbf{n}_1',\mathbf{m}_2\mathbf{n}_2}(\mathbf{q})=&\lambda^{-2} \prod_{i}^{x,y}N_{n_{1i}'}N_{m_{2i}}N_{n_{1'i}}N_{n_{2i}} \mathrm{e}^{-\frac{1}{2}\mathbf{q}^2\lambda^2}\sum_{u_i=0}^{\min[m_{1i}',m_{2i}]}\sum_{v_i=0}^{\min[n_{1i}',n_{2i}]}\binom{n_{2i}}{v_{i}}\binom{n_{1i}'}{v_{i}}\binom{m_{2i}}{u_{i}}\binom{m_{1i}'}{u_{i}}
    \notag\\
      &u_{i}!v_{i}!2^{u_i+v_i}(-1)^{n_{1i}'+m_{1i}'}
      \sqrt{\frac{\pi}{2}}\frac{H_{\sigma_{2i}+\sigma_{1i}'}(\lambda q_i/\sqrt{2})}{\sqrt{2^{\sigma_{2i}+\sigma_{1i}'}}} ,
\end{align}
where $\sigma_{1i}'=m_{1i}'+n_{1i}'-2u_i$, $\sigma_{2i}=m_{2i}+n_{2i}-2v_i$, and the normalization constant $N_n=\sqrt{\lambda/(\pi^{1/2}2^nn!)}$. Similarly,
\begin{align}
    h'_{\mathbf{m}_1'\mathbf{n}_1',\mathbf{m}_2\mathbf{n}_2}(\mathbf{q})=&\mathrm{e}^{-\frac{1}{4}\lambda^2q^2}\sum_{r_x=0}^{m_{1x}'}\sum_{r_y=0}^{m_{1y}'}\sum_{s_x=0}^{n_{2x}}\sum_{s_y=0}^{n_{2y}}(-1)^{m_{1x}'+m_{1y}'-r_x-r_y}\binom{n_{2x}}{s_x}\binom{n_{2y}}{s_y}\binom{m_{1x'}}{r_x}\binom{m_{1y}'}{r_y}\notag\\
    &\frac{N_{m_{1x}'}}{N_{r_x}}\frac{N_{m_{1y}'}}{N_{r_y}}\frac{N_{n_{2x}}}{N_{s_x}}\frac{N_{n_{2y}}}{N_{s_y}}(\lambda q_x)^{m_{1x}'+n_{2x}-s_x-r_x}(\lambda q_y)^{m_{1y}'+n_{2y}-s_y-r_y} h_{\mathbf{s}\mathbf{n}_1',\mathbf{m}_2\mathbf{r}}(\tfrac{1}{2}\mathbf{q}).
\end{align}
For the hole-hole exchange integral, we let $\begin{bmatrix}\mathbf{k}_1\\\mathbf{k}_2\end{bmatrix}=\begin{bmatrix}\frac{1}{2}\mathbf{k}+\mathbf{p}\\\frac{1}{2}\mathbf{k}-\mathbf{p}\end{bmatrix}$ and $\begin{bmatrix}\mathbf{k}_1'\\\mathbf{k}_2'\end{bmatrix}=\begin{bmatrix}\frac{1}{2}\mathbf{k}'+\mathbf{p}'\\\frac{1}{2}\mathbf{k}'-\mathbf{p}'\end{bmatrix}$. This gives
\begin{align}
    J^{\bm{m}\bm{m}'}_{\bm{n}\bm{n}'}=&
    \int d\mathbf{p}d\mathbf{k} d\mathbf{p}'d\mathbf{k}'
    W(\mathbf{k}-\mathbf{k}')
    g^{\bm{m}}_{\mathbf{p}+\frac{1}{2}\mathbf{k},\mathbf{p}-\frac{1}{2}\mathbf{k}}(\mathbf{k}-\mathbf{k}')\bar{g}^{\bm{m}'}_{\mathbf{p}'+\frac{1}{2}\mathbf{k}',\mathbf{p}'-\frac{1}{2}\mathbf{k}'}(\mathbf{k}-\mathbf{k}')
    \Upsilon^{\bm{n}\ast}_{\mathbf{p}+\frac{1}{2}\mathbf{k},\mathbf{p}-\frac{1}{2}\mathbf{k}}\Upsilon^{\bm{n}'\ast}_{\mathbf{p}'+\frac{1}{2}\mathbf{k}',\mathbf{p}'-\frac{1}{2}\mathbf{k}'}\notag\\
    =&
    4\int d\mathbf{q} d\mathbf{q}'
    W(\mathbf{q})\Delta^{\bm{m}\bm{n}}(\mathbf{q},\mathbf{q}')
    \Delta^{\bm{m}'\bm{n}'}(-\mathbf{q},\mathbf{q}') ,
\end{align}
where, in the second line, we change variable again with $\mathbf{q}=\mathbf{k}-\mathbf{k}'$ and $\mathbf{q}'=\mathbf{k}+\mathbf{k}'$ which gives
\begin{equation}
    \Delta^{\bm{m}\bm{n}}(\mathbf{q},\mathbf{q}')=\int d\mathbf{p}g^{\bm{m}}_{\mathbf{p}+\frac{1}{4}\mathbf{q}'+\frac{1}{4}\mathbf{q},\mathbf{p}-\frac{1}{4}\mathbf{q}'-\frac{1}{4}\mathbf{q}}(\mathbf{q})
     \Upsilon^{\bm{n}\ast}_{\mathbf{p}+\frac{1}{4}\mathbf{q}'+\frac{1}{4}\mathbf{q},\mathbf{p}-\frac{1}{4}\mathbf{q}'-\frac{1}{4}\mathbf{q}}.
\end{equation}
For $\lambda_1=\lambda_2$, we also have a simple exchange integral form, being
\begin{align}
    \Delta^{\bm{m}\bm{n}}(\mathbf{q},\mathbf{q}')
    =&
        (-1)^{n_{2i}+m_{2i}}[2\mathrm{e}^{-\frac{1}{4}\lambda^2q^2}\sum^{\mathbf{n}_1}_{\mathbf{s}_1=0}\sum^{\mathbf{m}_1}_{\mathbf{r}_1=0}\frac{N_{\mathbf{m}_1}}{N_{\mathbf{r}_1}}\frac{N_{\mathbf{n}_1}}{N_{\mathbf{s}_1}} (-1)^{n_{1i}-s_{1i}}\binom{n_{1i}}{s_{1i}}\binom{m_{1i}}{r_{1i}}(\tfrac{1}{2}q\lambda)^{\mathbf{m}_1+\mathbf{n}_1-\mathbf{r}_1-\mathbf{s}_1}h_{\mathbf{r}_1\mathbf{m}_2\mathbf{s}_1\mathbf{n}_2}(\mathbf{q}'-\tfrac{1}{2}\mathbf{q})\notag\\
    &-h_{\mathbf{m}_1\mathbf{m}_2\mathbf{n}_1\mathbf{n}_2}(\mathbf{q}')].
\end{align}
We can further integrating out $\mathbf{q}'$ analytical in $J^{\bm{m}\bm{m}'}_{\bm{n}\bm{n}'}$.

Therefore, the remaining problem is to integrate out $\mathbf{q}$. This integration can be done analytically for the Keldysh potential in Eq.~\eqref{eqn:W} as follows. First, we integrating out the angular dependence
\begin{equation}
    \int d\mathbf{q}W(\mathbf{q})(q_x\lambda)^{b_x}(q_y\lambda)^{b_y}\mathrm{e}^{-a q^2\lambda^2}=2\pi e^2\int dq\frac{q^{b_x+b_y}}{1+r_\ast q}\mathrm{e}^{-a q^2\lambda^2}\begin{cases}
    2B(\tfrac{b_x+1}{2},\tfrac{b_y+1}{2}),&b_x,b_y=\text{even} , \\
    0, &
    \end{cases}
\end{equation}
where $B(x,y)$ is the beta function. Further, integrating out $q$, we obtain
\begin{equation*}
    \int dq\frac{q^{b_x+b_y}}{1+r_\ast q}\mathrm{e}^{-a q^2\lambda^2}=
   \frac{\mathrm{e}^{-a \lambda^2/r_\ast^2}}{2\lambda}\Big(-\frac{\lambda }{r_\ast}\Big)^{b_x+b_y+1} [\pi \text{erfi}(\sqrt{a}\lambda/r_\ast)-\text{Ei}(a \lambda^2 /r_\ast^2)]
   -\sum_{j=0}^{b_x+b_y-1}(-\tfrac{\lambda}{r_\ast})^{b_x+b_y-j}\tfrac{1}{2}a^{-(1+j)/2}\Gamma(\tfrac{1+j}{2})
\end{equation*}
where erfi is the imaginary error function and Ei is the exponential-integral function.

We note that for large differences between electron's masses in conduction band, it is better to let $\lambda_1\neq\lambda_2$ and optimize those parameters separately. This can lead to better convergent rate for solving Wannier equation in Eq.~\eqref{eqn:T-WannierQ}, while making calculations significantly more complicated when evaluating nonlinearity numerically. It may not have useful analytical expression for in Eqs.~\eqref{eqn:I} and \eqref{eqn:J}. Instead, we may need to perform the $\mathbf{k}$-integration (8-dimensional) in Eq.~\eqref{eqn:nonlinearity} numerically.

\end{widetext}


\bibliography{trion}

\end{document}